\documentclass[reprint,showkeys,aps,pre,floats,twocolumn,superscriptaddress]{revtex4-1}
\usepackage[english]{babel}
\usepackage[utf8]{inputenc}
\usepackage{graphicx,epsfig}% Include figure files
\usepackage{graphics,dcolumn,bm,epic, eepic,fleqn,float}
\usepackage{amssymb,amsmath,amsfonts,multirow,rotate,color}
\usepackage{soul}

\usepackage[colorinlistoftodos, color=green!40, prependcaption]{todonotes}
\usepackage{amsthm}
\usepackage{mathtools}
\usepackage{physics}
\usepackage{xcolor}
\usepackage{hyperref}
\hypersetup{
    colorlinks=true,
    linkcolor={black!50!black},
    citecolor={blue!80!black},
    urlcolor={blue!80!black}
}
\usepackage{adjustbox}
\usepackage{placeins}
\usepackage{lipsum}
\usepackage{csquotes}

\makeatletter
\def\url@mystyle{%
  \@ifundefined{selectfont}{\def\UrlFont{\sf}}{\def\UrlFont{\tiny\ttfamily}}}
\makeatother
\definecolor{MyOrange}{rgb}{1,0.66,0.30}

\newcommand{\avg}[1]{\langle #1 \rangle}

\begin{document}

\title{Optimising the mitigation of epidemic spreading through targeted
  adoption of contact tracing apps}

\author{Aleix Bassolas}
\thanks{These authors contributed equally to
  this paper}
\affiliation{School of Mathematical Sciences, Queen Mary University
  of London, London E1 4NS, United Kingdom}

\author{Andrea Santoro}
\thanks{These authors contributed equally to
  this paper}
\affiliation{School of Mathematical Sciences, Queen Mary University
  of London, London E1 4NS, United Kingdom}

\author{Sandro Sousa}
  \affiliation{School of Mathematical Sciences, Queen Mary University
  of London, London E1 4NS, United Kingdom}

\author{Silvia Rognone}
\affiliation{School of Mathematical Sciences, Queen Mary University
  of London, London E1 4NS, United Kingdom}

\author{Vincenzo Nicosia}
\affiliation{School of Mathematical Sciences, Queen Mary University
  of London, London E1 4NS, United Kingdom}

\date{\today}

\begin{abstract}
The ongoing COVID-19 pandemic is the first epidemic in human history
in which digital contact-tracing has been deployed at a global
scale. Tracking and quarantining all the contacts of individuals who
test positive to a virus can help slowing-down an epidemic, but the
impact of contact-tracing is severely limited by the generally low
adoption of contact-tracing apps in the population. We derive here an
analytical expression for the effectiveness of contact-tracing app
installation strategies in a SIR model on a given contact graph. We
propose a decentralised heuristic to improve the effectiveness of
contact tracing under fixed adoption rates, which targets a set of
individuals to install contact-tracing apps, and can be easily
implemented. Simulations on a large number of real-world contact
networks confirm that this heuristic represents a feasible alternative
to the current state of the art.
\end{abstract}


\maketitle

Since the first human infection towards the end of 2019, the spread of
the SARS-COV-2 virus has caused an unprecedented shock around the
world, with serious repercussions in all aspects of our social and
economic
activities~\cite{mckibbin2020economic,mckibbin2020global,holman2020unfolding},
and a number of casualties that has already passed the two millions
figure and is unfortunately due to rise further in the near
future~\cite{mega2020covid}. The initial efforts to curb the spread of
the disease focused on non-pharmaceutical interventions including
travel bans, lockdowns, and curfews\cite{Perra2021}. These measures
are able to drastically reduce the opportunities of contacts between
infected and susceptible people and thus the spread of a
virus~\cite{balcan2009,riley2007,ferguson2005,colizza2007,PastorSatorras2015},
but also have non-negligible effects on the economy and social
life~\cite{flaxman2020estimating,cowling2020impact,peak2017comparing}. After
the first wave of infections in February-May 2020 and thanks to a
better understanding of the specific transmission dynamics of
SARS-COV-2~\cite{arenas2020,chang2020mobility,di2020impact,wiersinga2020pathophysiology,haug2020,aguilar2020impact},
many countries have implemented some sort of ``test-trace-treat''
system based on digital contact tracing
\cite{rodriguez2021population,bradshaw2021bidirectional}.  Some of
these systems consist on deploying contact tracing (CT) apps on mobile
phones which allow to identify and isolate individuals who have been
in contact with infected ones, thus disrupting secondary infections
paths as early as possible. With contact tracing in place, many
countries have been able to partially re-open several sectors of their
economy and to diminish the damage of prolonged
disruptions~\cite{aleta2020,panovska2020determining,cohen2020countries,abueg2020modeling,jiang2020survey,kerr2020controlling,kucharski2020effectiveness,smith2020adherence}

An effective digital contact tracing strategy should aim at maximising
the probability of detecting contacts between infected and susceptible
individuals, and it would completely eradicate contagion in the ideal
case where CT apps are installed by the totality of a population
\cite{eames2003contact,klinkenberg2006effectiveness,kretzschmar2020impact,ferretti2020quantifying,keeling2020efficacy,park2020contact,barrat2020CT,gardenes2020spreadingCT,kryven2020CT,bianconi2020messageapps}.
However, throughout the SARS-COV-2 pandemic the percentage of the
population with CT apps installed has remained quite low, between 5\%
and 20\% in most countries~\cite{StatistaCTPerc}, resulting in a
dramatically decreased efficiency of contact-tracing.

Here we focus on the problem of determining the set of nodes which
should install CT apps in order to optimise the effect of contact
tracing, i.e., to maximally slow-down spreading and reduce the
incidence of a disease, under the assumption that the rate of CT app
adoption is fixed. We provide an analytic derivation to quantify the
decrease of the basic reproduction number caused by a generic CT
installation strategy, and we show that uniform random installation --
which is the strategy implicitly adopted by governments when people
are simply asked to install a CT app -- has the worst performance of
all. We find that relatively simple targeting strategies based on the
structure of the contact network are significantly more efficient in
reducing the number of secondary infections at low adoption rates, in
both synthetic and real-world systems.

\begin{figure}[!t]
  \begin{center}
    \includegraphics[width=3.1in]{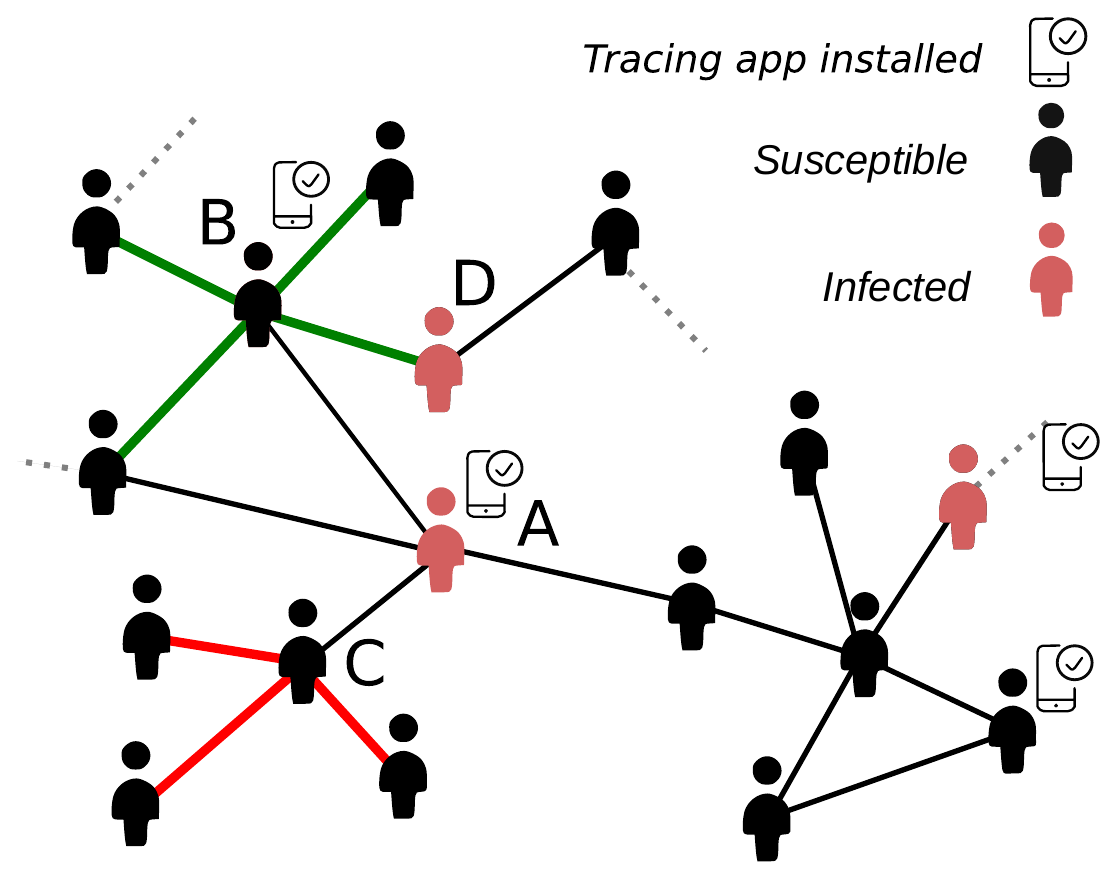}
  \end{center}
  \caption{\textbf{Effect of contact-tracing apps on secondary
      infections on a contact network.}  Contact tracing apps can only
    detect potential contagious contacts if both the infected and the
    susceptible individual have a CT app installed. In this case, A
    can in principle infect any of his direct neighbours, including B
    who has a CT app installed as well, and the CT app cannot do
    anything to avoid this. However, if the CT system detects a
    contact between A and B and then A tests positive, then B can be
    contacted and put into quarantine, thus disrupting all the
    potential infection paths to the direct contacts of B (solid green
    lines). If C has a contact with A, instead, C has no way of
    knowing whether the contact resulted in an infection or not, he
    will not be notified when A tests positive, and he will not go
    into quarantine. In this case, all the neighbours of C are at risk
    of catching the disease (solid red lines). Similarly, the fact
    that B has a CT app installed cannot safeguard her from being
    infected by D (who does not have a CT app installed) and passing
    the infection to her neighbours while she has no symptoms.}
  \label{fig:fig1}
\end{figure}

\section*{Results}
In Fig.~\ref{fig:fig1} we report a sketch of a fictious contact
network, where some individuals are infected (pink), some other are
susceptible (black), and some have a CT app installed (indicated by
the mobile icon). A perfect lockdown would remove almost all the links
in that graph (with the only exception of those among people belonging
to the same household), so that infected individuals will eventually
be unable to find any susceptible person to pass the disease on. When
only contact tracing is in place, instead, some infections are still
unavoidable, either due to a limited app adoption rate or to a delay
in the notification of test results and in isolation of subjects
exposed to infected ones. As made evident by Fig.~\ref{fig:fig1},
maximising the impact of contact tracing corresponds to maximising the
probability that the potential transmission of the disease between two
individuals is detected, since only contacts among people with CT apps
installed can be detected and traced back. This intuitively
corresponds to maximising the number of edges among the individuals
with CT app installed, i.e., the density of the subgraph induced by
the nodes with CT apps, under the constraint that only a fraction $r$
of the population will have the CT app installed.

\begin{figure*}[!t]
    \begin{center}
        \includegraphics[width=6.2in]{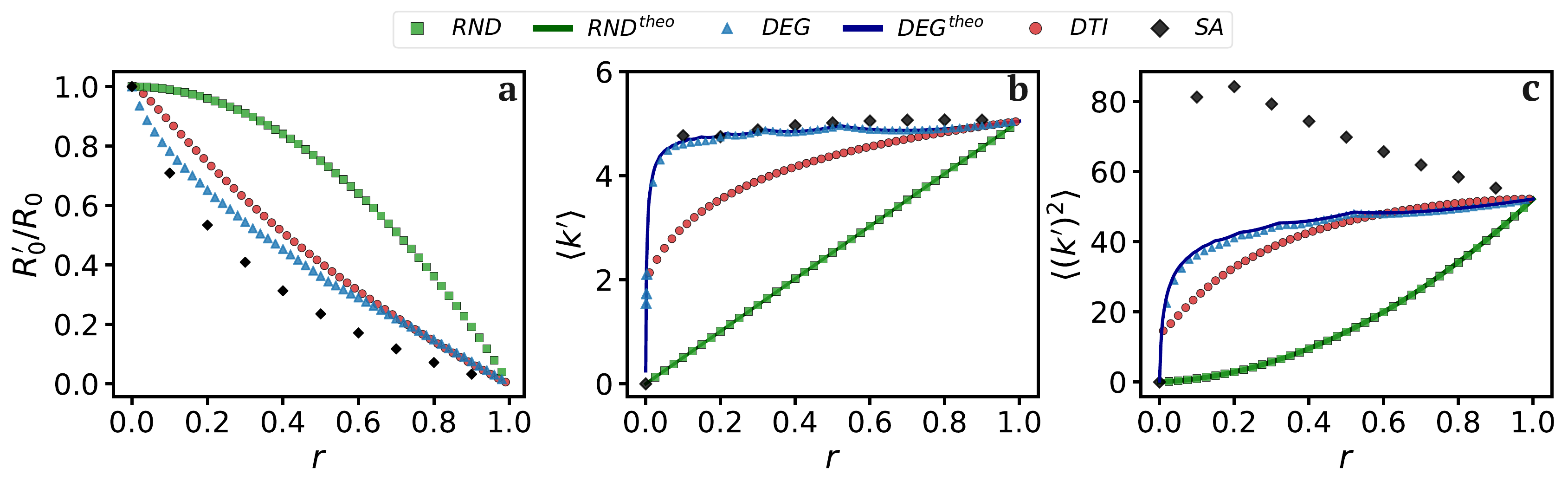}
    \end{center}
    \caption{\textbf{Effect of different CT strategies on the induced
        subgraph as a function of app adoption rate $r$.} (a) The
      ratio ${R'_0}/{R_0}$ as a function of the CT adoption rate $r$
      for the Random (RND), degree-based (DEG), distributed targeting
      (DTI), and Simulated Annealing (SA) strategies, calculated using
      Eq.~(\ref{eq:R_0_theo}). (b) Average degree of $G'$ as a
      function of the adoption rate $r$ for each of the strategies.
      The theoretical predictions for the RND and DEG strategies
      (solid lines) were obtained using Eq.~(\ref{eq:k_avg_random})
      and Eq.~(\ref{eq:k_avg_degree}), respectively. (c) Second moment
      of the degree distribution of $G'$ as a function of the adoption
      rate $r$ for each of the strategies. The plots correspond to an
      ensemble of configuration model graphs with degree distribution
      $P(k)\sim k^{-3}$ and $N=10^4$ nodes. Results averaged over 100
      realisations.}
    \label{fig:fig2}
\end{figure*}

\vspace{0.2cm}
\noindent
\textbf{Reduction of $R_0$ in a SIR+CT dynamics.} We consider here a
SIR+CT model, which is a classical Susceptible-Infected-Recovered (SIR)
model on a static contact graph~\cite{PastorSatorras2015}, with the
addition of ideal contact tracing. This means that any susceptible
node with installed CT app is quarantined (recovered) as soon as one
of their contacts with CT app installed gets infected. The parameters
of the model are the probability $\beta$ that an infected individual
passes the disease to each of its susceptible neighbours, and the
probability $\mu$ that an infected individual is removed (due to
either recovery or death). We call the contact graph $G(V,E)$, with
$N=|V|$ nodes and $K=|E|$ edges, and we denote by $G'(V', E')$ the
subgraph of $G$ induced by CT app installations, i.e., such that $V'$
is the set of nodes in $G$ with CT apps and $E'$ is the set of edges
among nodes in $V'$. We quantify the effect of the installation of CT
apps in a certain subset $V'$ of nodes by computing the reduction of
the basic reproduction number $R_0$, that is the expected number of
secondary infections caused by a single contagion event. Let us assume
that the generic node $\ell$ is infected and has passed the disease to
its neighbour $i$. The expected number $R_i$ of secondary infections
caused by $i$ while it remains infected depends on whether $i$ is in
$V'$, and on how many of its $k_i$ neighbours are in $V'$ as well. In
particular, if $i\notin V'$, $R_i = \frac{\beta}{\mu}(k_i - 1)$ as in
the classical SIR (we have to remove $\ell$ from the count, hence the
$k_i - 1$)~\cite{PastorSatorras2015}. If $i\in V'$, instead, there are
two possible cases: \textit{a)} if $\ell\notin V'$, the contact
between $\ell$ and $i$ remains undetected, and $i$ can potentially
infect $R_i=\frac{\beta}{\mu}(k_i-1)$ more nodes, as in the classical
SIR. \textit{b)} If instead $\ell\in V'$ as well, then the contact
with $i$ gets detected by the CT system and $i$ goes into
self-isolation immediately, thus avoiding any secondary infection. If
we denote by $k'_i$ the degree of node $i$ in $G'$, the expected
number of infections potentially caused by the infection of $i$ is
equal to
\begin{equation*}
  R_{i} = \frac{\beta}{\mu}(k_i-1) \frac{k_i-k_i'}{k_i}
\end{equation*}
and the average number of secondary infections potentially caused by
each node infected by $\ell$ is given by:
\begin{equation*}
  R_{\ell} = \frac{\beta}{\mu}\frac{1}{k_{\ell}}\sum_{i}a_{\ell i}(k_i-1) \frac{k_i-k_i'}{k_i}
\end{equation*}
where $a_{\ell i}$ are the entries of the adjacency matrix of the
contact graph $G$. By averaging $R_{\ell}$ over all the nodes of $G$
we obtain the value of the basic reproduction number in presence of
contact tracing (see Methods for details):
\begin{equation}
  \begin{array}{rl}
    R'_{0} & = R_{0} - \frac{1}{N}\frac{\beta}{\mu}
  \sum_{\ell}\frac{1}{k_{\ell}}\sum_{i}a_{\ell i} \frac{k'_i}{k_i}(k_i
  - 1)
  \end{array}
  \label{eq:R_0_theo}
\end{equation}
where $R_0$ is the basic reproduction number of the classical SIR
dynamics on $G$~\cite{PastorSatorras2015}. As made clear by
Eq.~(\ref{eq:R_0_theo}), we can minimise the value of $R_{0}'$ by
using a generic optimisation algorithm to compute:
\begin{equation}
  \max_{G'} \quad \mathcal{F}(G') =
  \sum_{\ell}\frac{1}{k_{\ell}}\sum_{i}a_{\ell i} \frac{k'_i}{k_i}(k_i
  - 1)
  \label{eq:cost_function}
\end{equation}
over the ensemble of possible choices of $G'$. Notice that if the
entire population installs CT apps (i.e., if $k'_i = k_i\>\forall i\in
V$) we trivially get $R'_0 = 0$ (see Methods and Supplementary Note 1
for details).

\vspace{0.2cm}
\noindent
\textbf{CT targeting strategies.} If we assume that we can install the
contact tracing app only to a fraction $r\in[0,1]$ of ``willing''
individuals, Eq.~(\ref{eq:cost_function}) states that a good CT
installation strategy should include in $G'$ nodes having a high
degree in $G$ (so that the ratio $\frac{(k_i-1)}{k_i}$ is as large as
possible), and, at the same time, a high number of connections to
other nodes in $G'$ (i.e., so that $k'_i$ is as large as
possible). The most basic strategy to select a fraction $r$ of the $N$
individuals to install CT apps consists in asking the population to
install a CT app in their mobile phones, under the assumption that
each individual will comply with probability equal to $r$,
irrespective of any of their specific social or behavioural
characteristics. In this case, the total number of installations will
be distributed according to a Binomial, with mean equal to $rN$. In
the following, we call this strategy ``uniform random installation''
(RND).

A second strategy consists in explicitly targeting all the potential
\textit{super-spreaders}
\cite{gomez2020mapping,miller2020transmission}. In practice, we ask
the $rN$ individuals with the largest number of contacts (links) in
$G$ to install the app, assuming that they will all comply with
probability $1$. This strategy is indeed utopistic since it requires
full knowledge of the contact network and full compliance by the
selected nodes. In the following, we call this strategy ``degree-based
installation'' (DEG).

Here we propose and study a constructive strategy to maximise
Eq.~(\ref{eq:cost_function}) that does not require detailed global
information on $G$, and is thus amenable to a distributed
implementation. We start from a CT set that contains only the node
with the largest degree in $G$. Then, at each subsequent step $t$ we
add to the CT set one of the neighbours $i$ of any of the nodes in
$V'$, with probability proportional to the total number of neighbours
of that node that are already in $V'$ (see Methods for details). This
creates a ``social pressure'' on individuals with no CT app installed
which is proportional to the number of their contacts already in
$V'$. We call this strategy ``distributed targeting installation''
(DTI).

\begin{figure*}[!htbp]
    \begin{center}
        \includegraphics[width=6.2in]{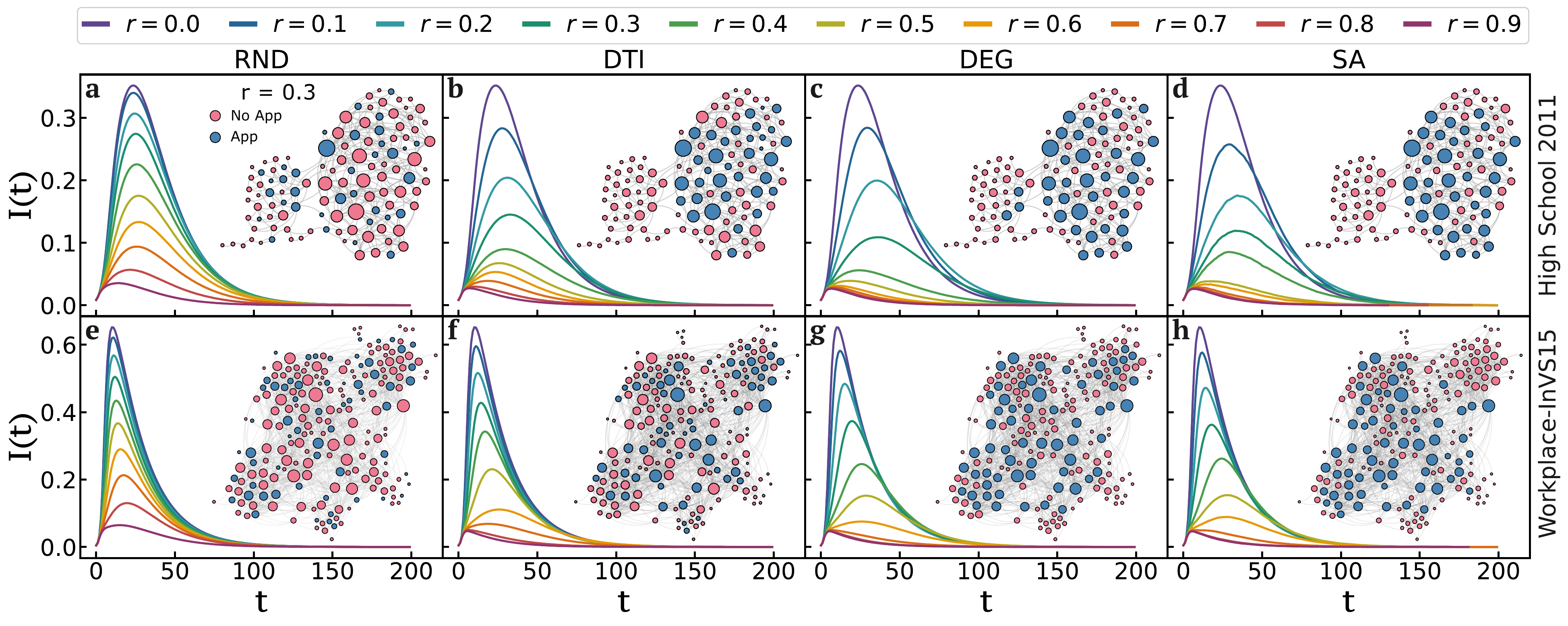}
    \end{center}
    \caption{\textbf{Impact of CT strategy and adoption rate on the
        epidemic peak of SIR+CT in real-world networks}. The evolution
      of the disease in a SIR+CT model (here for $\beta=0.1$, $\mu =
      0.05$) depends heavily on the adoption rate $r$ and on the
      strategy used to select which individuals will have a CT app
      installed. We show here the results on two real-world social
      networks, namely, the high-resolution face-to-face contact data
      respectively recorded in a high school (a-d) and a workplace
      (e-h). We have applied a threshold to both contact networks:
      keeping only contacts larger than 240 seconds for the High
      School and 10\% of the links with the largest weight for the
      Workplace (See Methods for details).  At adoption rates $r
      \gtrsim 0.4$, the RND strategy (a, e) displays a higher
      percentage of infected compared to DTI (b, f), DEG (c, g) and SA
      (d, h). These differences are likely linked to the structure of
      the subgraph induced by each CT strategy. Typical examples of
      those graphs for each strategy and $r=0.3$ are shown in the
      insets, where the size of each node is proportional to its
      degree and nodes with CT app installed are indicated in blue.}
    \label{fig:fig3}
\end{figure*}

In Fig.~\ref{fig:fig2}(a) we plot the ratio $\frac{R'_0}{R_0}$ as a
function of the CT adoption rate $r$ for the RND, DEG, and DTI
strategies, on an ensemble of configuration model graphs with
power-law degree distributions. As a reference, we also report the
results obtained by optimising Eq.~(\ref{eq:cost_function}) by means
of Simulated Annealing (SA). It is worth noting that RND is the
worst-performing strategy overall, characterised by a much slower
decrease of $R'_0$ with $r$. Conversely, degree-based installation is
close to the theoretical limit established by SA, and produces a
noticeable decrease of $R_0$ already for quite small values of
$r$. Remarkably, the performance of DTI is quite close to that of DEG,
although DTI is not using any global information about the structure
of $G$. In Fig.~\ref{fig:fig2}(b-c) we show how the first and second
moment of the degree distribution $\widetilde{P}(k')$ of the subgraph
$G'$ vary with $r$ for each of the four strategies. More details on
the derivation of the full degree distribution of $G'$ in RND and DEG
are reported in Methods, while Supplementary Fig. 1 shows the perfect
agreement between the empirical and analytical degree distributions
for these two strategies.

It is worth noting that under the DEG strategy $\avg{k'}$ increases
very sharply with $r$ and is already quite similar to the value of
$\avg{k}$ in $G$ for very small values of $r$. On the other hand, in
RND $\avg{k'}$ increases only linearly with $r$ (see Methods for
details), while the performance of DTI is in between those
two. However, these plots make it clear that the sheer density of $G'$
is not the only important ingredient for CT app installation.  Indeed,
SA can attain consistently lower values of $\frac{R'_0}{R_0}$ than
DEG, although the values of $\avg{k'}$ produced by SA are almost
identical to that provided by DEG (see Fig.~\ref{fig:fig2}b).

\noindent
\textbf{SIR+CT in real-world graphs.} In Fig.~\ref{fig:fig3} we show
the ratio of infected nodes $I(t)$ for the four strategies with
different values of $r$ on two real-world contact networks,
respectively the network of friendship in a high school (top panels)
and at a workplace (bottom panels)\cite{Sociopatterns}. In the inset
of each panel we report a typical composition of $V'$ for each
strategy. It is evident that, even at low adoption rates, the DEG and
DTI strategy can heavily mitigate the incidence of the disease better
than RND. This is most probably because DEG and DTI are
targeting different sets of nodes than RND, and in general end up
selecting nodes with high degree which result in a higher edge density
in $G'$.

\begin{figure*}[!tbp]
  \begin{center}
    \includegraphics[width=6.2in]{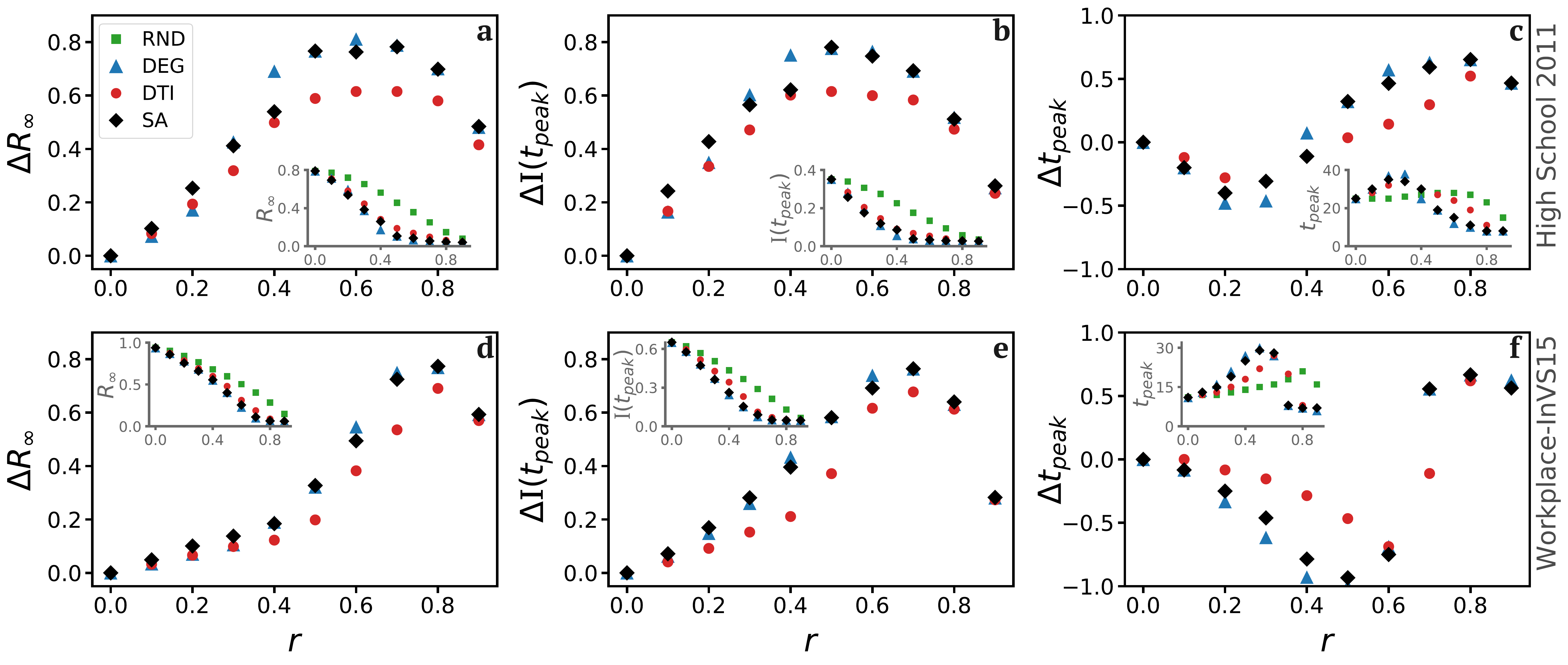}
  \end{center}
  \caption{\textbf{Comparison of epidemic indicators under different
      CT strategies}. Relative decrease with respect to random
    installations of the total number of recovered $\Delta R^{s}_{\infty}$,
    height of the infection peak $\Delta I^{s}(t_{\rm peak})$ and position of the
    peak $\Delta t^{s}_{\rm peak}$ (see Eq. (\ref{eq:indicators}))
    for DTI, DEG, and SA targeted installation, in
    the same contact networks shown in Fig.~\ref{fig:fig3}. The inset of
    each panel reports the plot of the raw variable, respectively
    $R_{\infty}$ (panel a and d), $I(t_{\rm peak})$ (panel b and e)
    and $t_{\rm peak}$ (panel c and f).}
  \label{fig:fig4}
\end{figure*}

The dynamics of SIR+CT in the two systems exhibits some noticeable
qualitative differences when distinct strategies are adopted, with
respect to the height of the infection peak (the maximum incidence of
the disease), the actual position of the peak (the time at which it
occurs), and the overall duration of the epidemic. Interestingly, the
position of the peak shifts to the right (delays) at small values of
$r$ for the DEG, DTI and SA strategies. Conversely, the peak starts to
recede (it is anticipated) with respect to the baseline when $r$
becomes larger than a certain threshold, which depends on the
particular structure of the contact network. While low values of $r$
lead to a delay in the dynamics -- the peak shifts to the right --
large enough values of $r$ effectively break the network into a number
of disconnected components, resulting in a considerable disruption of
the spreading -- shift to the left.

To better understand these qualitative differences, we look at three
key properties of the epidemic curve, namely the total number of
individuals recovered $R_{\infty}$, the maximum number of individuals
infected across the duration of the epidemic $I(t_{\rm peak})$ and the
time to reach the infection peak $t_{\rm peak}$. In particular, we
compute the relative performance of each strategy $s$ (being it either
DEG, DTI, and SA) with respect to RND using the quantities:
\begin{equation}
  \begin{array}{rl}
    \Delta R^{s}_{\infty} & = 1 - \frac{R^s_{\infty}}
    {R^{\rm RND}_{\infty}} \\
    \Delta I^{s}(t_{\rm peak}) & = 1 - \frac{I^s_{t_{\rm
    peak}}}{I^{\rm RND}(t_{\rm peak)}} \\
    \Delta t^{s}_{\rm peak} & = 1 - \frac{t^{s}_{\rm peak}}
    {t^{\rm RND}_{\rm peak}}
  \end{array}
  \label{eq:indicators}
\end{equation}
The results are shown in Fig.~\ref{fig:fig4}. We found that in the
high school network an adoption rate of $r=0.2$ can decrease the
number of infected individuals at the peak by as much as 40\%
(Fig.~\ref{fig:fig4}b) for all the three strategies. At the same
adoption rate, the total number of infected individuals decreases by
20\% (Fig.~\ref{fig:fig4}a) and the peak is delayed by approximately
40\% (Fig.~\ref{fig:fig4}c). For $r=0.5$ we witness a substantially
stronger mitigation, where the peak is reduced by up to 70\% and the
total number of infected is reduced by up to 80\%. The effectiveness
of targeting strategies is somehow less pronounced in the workplace
graph at similar adoption rates (bottom panels).  Interestingly, in
both networks the DTI strategy performs similarly to DEG and SA, as
can be observed in the insets in Fig.~\ref{fig:fig4}, which show the
raw values of $R_{\infty}$, $I(t_{\rm peak})$ and $t_{\rm peak}$.

\begin{figure*}[!t]
    \begin{center}
        \includegraphics[width=6.2in]{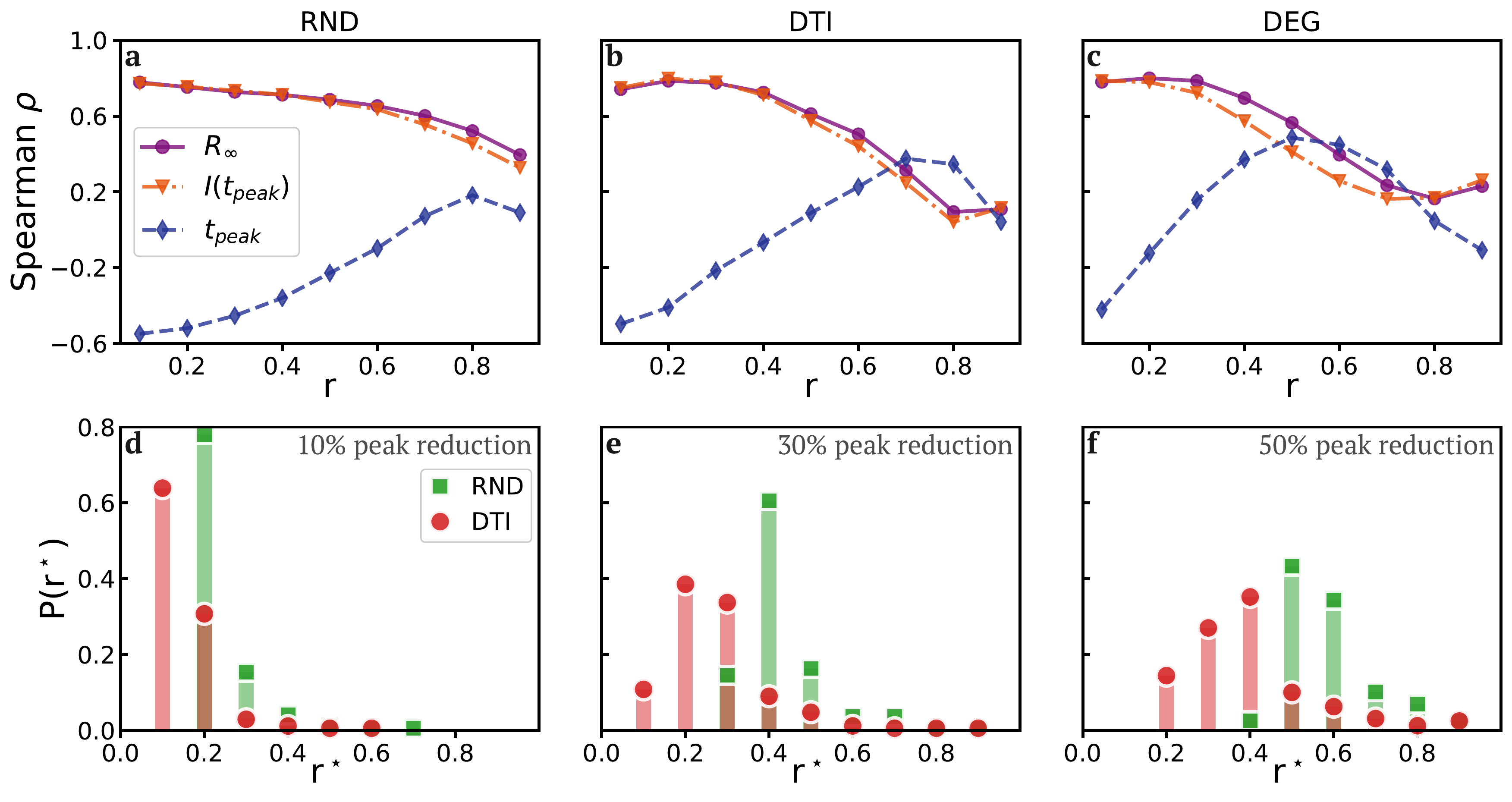}
    \end{center}
    \caption{\textbf{Correlations with network structural measures and
        performance of CT strategies to mitigate an epidemic}. For the
      168 real-world contact networks analysed, panels (a-c) report as
      a function of $r$ the Spearman rank correlation between the
      analytical value of $R'_{0}$ (Eq. (\ref{eq:R_0_theo})) and the
      epidemiological indicators $R^{s}_{\infty}$ (purple),
      $I^{s}(t_{\rm peak})$ (orange) and $t^{s}_{\rm peak}$ (blue) for
      the RND (a), DEG (b) and DTI (c) strategies. Panels (d-f) show
      the distribution of minimum adoption ratios $r^{*}$ needed to
      produce a $10\%$ (d), $30\%$ (e) and $50\%$ (f) for the DTI
      (red) and RND (green) strategies. Overall, DTI largely
      outperforms RND, which is the strategy currently adopted by many
      governments.}
    \label{fig:fig5}
\end{figure*}

We simulated SIR+CT in 84 unique real-world contact network data sets,
filtered by applying two different thresholds, for a total of 168
undirected
graphs~\cite{Starnini2013,vanhems2013,Barrat2014,Sociopatterns,
  Genois2015,Genois2018,gadar2020multilayer,sapiezynski2019interaction}
(See Methods for details).  In Fig.~\ref{fig:fig5}(a-c), we report for
the RND, DEG and DTI strategies, and several adoption rates $r$, the
spearman correlation between the analytical $R'_{0}$ and the
epidemiological indicators $R_{\infty}$, $I(t_{\rm peak})$ and $t_{\rm
  peak}$. The correlation with both $R_{\infty}$ and $I(t_{\rm peak})$
is high for the three strategies confirming the analytical predictions
of Eq. (\ref{eq:R_0_theo}) despite the small size of the graphs and
the presence of degree-degree correlations. Still, as $r$ increases we
observe a decrease in the correlations, likely due to finite size
effects. The correlation between $R'_0$ and $t_{\rm peak}$ displays a
much richer behaviour: we start with a significant but negative
correlation for small $r$, which changes sign until it reaches a
maximum. We conjecture that the change of sign is related to the
movement of the peak: while in the small-$r$ regime lower values of
$R'_{0}$ contribute to a delay of the peak, for larger values of $r$
we observe a stronger anticipation of the peak.  The value $r_{\rm
  peak}$ at which the correlation peaks depends on the strategy in
use, the more efficient it is, the lower is the value of $r_{\rm
  peak}$.  The concrete value of $r_{\rm peak}$ seems thus related to
a certain structural cutoff of the graphs.

In a realistic scenario, in which the adoption rate is not fixed but
needs to be promoted, we might be more interested in the minimum
adoption rate $r^{*}$ needed on each network to obtain a given
reduction of the infection peak with respect to the absence of contact
tracing. In Fig.~\ref{fig:fig5}(d-f) we report the histograms of the
value of $r^*$ in DTI and RND across the 168 networks when we set a
reduction in the peak $I(t_{\rm peak})$ of 10\%, 30\% and 50\%,
respectively.  We found that DTI can achieve a reduction of 30\% of
the peak in 85\% of the networks with an adoption rate smaller than
$0.3$ (panel e), while the RND strategy would need an adoption rate of
$0.5$ to achieve an equivalent reduction.

While the strategies analysed here require some level of global
information and full compliance by individuals to install the app, we
have obtained qualitatively similar results with other decentralised
strategies based on local information, and with a tunable level of
compliance (see Supplementary Note 2 and Supplementary Figs. 2-7 for
details). Moreover, qualitatively similar results are also obtained
when considering a SIR+CT with maximum delay, i.e., where an
individual with the CT app installed goes into quarantine only when it
becomes infected (see Supplementary Note 3 and Supplementary
Figs. 7-9).

\vspace{0.2cm}
\noindent
\textbf{Conclusion.}  We have shown here that the random CT app
installation -- considered in recent studies on the
topic~\cite{barrat2020CT,gardenes2020spreadingCT,kryven2020CT} and
adopted by many governments -- is the less effective strategy to
mitigate the effects of a pandemic through contact tracing. The
theoretical argument presented here, which links the reduction of
$R_0$ to the structure of the subgraph $G'$ induced by CT, holds for
any graph under any CT strategy. In particular, Eq.~\ref{eq:R_0_theo}
provides a concrete recipe to maximise the effectiveness of
a CT app deployment as we have shown in real-world systems.

The reduction of a disease incidence attainable by the DTI strategy is
comparable with degree-based targeting, which performs similarly to
the optimal targeting obtained through simulated annealing. A notable
advantage of using DTI over DEG is that it does not require any global
information about the graph $G$ and it can be implemented on a
distributed manner. For instance, one could ask every new individual
who installs the CT app to broadcast a message to all his contacts,
asking them to install it as well.  By doing so, each contact with no
app installed will be subject to a level of ``social pressure''
linearly proportional to the number of contacts who already have the
CT app installed (in agreement with the heuristic algorithm of which
DTI is based), consequently increasing the likelihood that he will
also install it.

Although several effective vaccines have been made available
recently~\cite{jeyanathan2020immunological}, mass vaccination
campaigns are at their initial stage in many countries and might last
for several months before a sufficient percentage of the population is
vaccinated. Moreover, SARS-COV-2 variants may develop vaccine
resistance and prolong the duration of the epidemic, creating an
unsustainable loop of vaccine updates and vaccination campaigns.
Hence, reducing the spreading of the virus by detecting potential
infected individuals and limiting their contacts -- through digital
contact tracing -- is still essential~\cite{Gozzi2021importance}.  The
results shown here suggest that governments could significantly
improve the effectiveness of contact tracing programs by implementing
targeting CT app installations, not only for the ongoing COVID-19 but
any major epidemic event in the future.

\section*{Methods}

\subsection*{Distributed Targeting Installation strategy}

The proposed heuristic constructive algorithm to optimise
Eq.~(\ref{eq:R_0_theo}), denominated DTI, starts with a set
$V'(t=0)$ containing the node of largest degree in $G$. At
each step $t$, we consider the set $S(t)$ of nodes which have
at least one neighbour in $V'(t)$, then, a node $i$ is selected
at $t+1$ from $S(t)$ and added to $V'(t)$ with probability:
\begin{equation}
  P(i; t) = \frac{\sum{j\in V'(t)}a_{ij}}{\sum_{i\in S(t)}\sum_{j\in V'(t)}a_{ij}}
\end{equation}
i.e., node $i$ is selected linearly proportional to the number
of neighbours it has in $V'(t)$.

\subsection*{Properties of the subgraph induced by app installation}

We provide here a sketch of the derivation of the first two moments of
the degree distribution of the subgraph $G'$ obtained from a graph $G$
by considering only the nodes which have the contact-tracing app
installed and the edges among them. The full derivations are provided
in Supplementary Note 1.  In the case of random
installation strategy, the probability that a node installs the
contact tracing app is uniform across all nodes. As a consequence, the
probability that a node with degree $k$ in $G$ has degree $k'$ in $G'$
is given by the Binomial distribution:
\begin{equation}
  Prob(k'_i = k'|k_i = k) = \binom{k}{k'} r^{k'}(1-r)^{k-k'}.
  \label{eq:P_kprime_k}
\end{equation}
This means that the expected degree in $G'$ of a node that has degree
$k$ in $G$ is just:
\begin{equation}
  E\left[k_i'\right] = rk_i
  \label{eq:expected_k_prime_rnd}
\end{equation}

The degree distribution of the subgraph $G'$ can be obtained by
summing the probability in Eq.~(\ref{eq:P_kprime_k}) over all possible
values of $k$, from which we obtain:
\begin{equation*}
  \widetilde{P}_{\rm RND}(k') = \sum_{k=0}^{N-1}\binom{k}{k'}
  r^{k'}(1-r)^{k-k'}
\end{equation*}
Finally, for the first two moments of $\widetilde{P}_{\rm RND}(k')$
we get:
\begin{equation}
  \avg{k}_{\rm RND} = \sum_{j=0}^{N-1}j\widetilde{P}_{\rm RND}(j) = r\avg{k}_{G}
    \label{eq:k_avg_random}
\end{equation}
and:
\begin{equation}
  \avg{k^2}_{\rm RND} = \sum_{j=0}^{N-1}j^2\widetilde{P}_{\rm RND}(j) = r^2\avg{k^2}_G+r(1-r)\avg{k}_G.
   \label{eq:k2_avg_random}
\end{equation}

To compute the degree distribution of $G'$ for degree-based
installations, we start from the observation that a node is in $G'$
only if its degree is $k_i \ge \widetilde{k}$, where $\widetilde{k}$
is obtained by solving the inequality:
\begin{equation}
  \sum_{\widetilde{k}}^{N}P(k) \ge r
\end{equation}
where $P(k)$ is the degree distribution of $G$. Now, the probability
that one of the $k_i$ neighbours of $i$ is in $G'$ is equal to:
\begin{equation}
  Q_{\widetilde{k}}(i) = \sum_{k=\widetilde{k}}^{N-1} P(k|k_i)
\end{equation}
where $P(k|k_i)$ is the conditional probability of finding in $G$ a
node with degree $k$ by following one of the edges of a node with
degree $k_i$, chosen uniformly at random. In the special case of
graphs with no degree-degree correlations, $P(k|k') =
\frac{kP(k)}{\avg{k}}= q_k$, so we have:
\begin{equation}
  Q_{\widetilde{k}}(i) = \sum_{k=\widetilde{k}}^{N-1} q_k = \widetilde{r}
  \quad \forall i
  \label{eq:r_tilde}
\end{equation}
In the absence of degree-degree correlations, the probability of any
two nodes to be connected does not depend on their degree, by
definition. Hence, the probability that a node of $G'$ has a degree
equal to $k'$ is given again by the Binomial distribution:
\begin{equation}
  Prob(k'_i = k' | k_i = k) =
  \binom{k}{k'}\widetilde{r}^{k'}(1-\widetilde{r})^{k-k'}, \quad k\geq
  \widetilde{k}
\end{equation}
while $Prob(k'_i=k' | k_i = k)=0$ if $k<\widetilde{k}$. In particular,
this means that the expected value $E\left[k_i'\right]$ is equal to:
\begin{equation}
  E\left[k_i'\right] = \widetilde{r}k_i
  \label{eq:expected_k_prime_deg}
\end{equation}
Notice that $\widetilde{r}$ has the same role that $r$ has in the
equations for uniform random installation. With an argument in all
similar to that used for random installation, we obtain:
\begin{equation}
  \widetilde{P}(k') = \sum_{k=\widetilde{k}}^{N-1}P(k)
  \binom{k}{k'}\widetilde{r}^{k'}(1-\widetilde{r})^{k-k'}
  \label{eq:Pk_inst_degbased}
\end{equation}
Here $\widetilde{P}(k')$ represents the probability to find a node of
$G$ which has degree $k'$ in the subgraph induced by app
installations. To obtain the actual degree distribution in the induced
subgraph, i.e., the probability that one of the nodes of $G'$ has
degree $k'$, we must rescale $\widetilde{P}(k')$ to the nodes in $G'$,
i.e., we consider the probability distribution:
\begin{equation}
  \widetilde{P}_{\rm DEG}(k') = \frac{1}{r}\widetilde{P}(k')
\end{equation}
It is easy to show that $\widetilde{P}_{\rm DEG}(k')$ is correctly
normalised:
\begin{equation}
  \begin{array}{rl}
    \displaystyle{\sum_{k'=0}^{rN-1}\widetilde{P}_{\rm DEG}(k')} & =
      \displaystyle{\frac{1}{r}\sum_{k'=0}^{rN-1}\sum_{k=\widetilde{k}}^{N-1}
      P(k)\binom{k}{k'}\widetilde{r}^{k'}(1-\widetilde{r})^{k-k'}}\\
    & =\displaystyle{\frac{1}{r}\sum_{k=\widetilde{k}}^{N-1}P(k)\sum_{k'=0}^{rN-1}\binom{k}{k'}\widetilde{r}^{k'}(1-\widetilde{r})^{k-k'}}\\
    & = \displaystyle{\frac{1}{r}\sum_{k=\widetilde{k}}^{N-1}P(k) = 1}
  \end{array}
\end{equation}

The average degree in the induced graph is obtained as follows:
\begin{equation}
  \begin{array}{rl}
    \avg{k}_{\rm DEG} & =
    \displaystyle{\sum_{k'=0}^{rN-1}k'\widetilde{P}_{\rm DEG}(k')}\\
      & = \displaystyle{
    \frac{1}{r}\sum_{k=\widetilde{k}}^{N-1}P(k)\sum_{k'=0}^{rN-1}k'\binom{k}{k'}\widetilde{r}^{k'}(1-\widetilde{r})^{k-k'}}\\
    & = \displaystyle{\frac{1}{r}\sum_{k=\widetilde{k}}^{N-1}P(k)k\widetilde{r} =
    \frac{\widetilde{r}}{r}\sum_{k=\widetilde{k}}^{N-1}kP(k) = \frac{\widetilde{r}^2}{r}\avg{k}}
  \end{array}
  \label{eq:k_avg_degree}
\end{equation}
where we have used the fact that $\sum_{k=\widetilde{k}}^{N-1}kP(k) =
\widetilde{r}\avg{k}$ as per the definition of $\widetilde{r}$ in
Eq.~(\ref{eq:r_tilde}). Similarly, for the second
moment we obtain:
\begin{equation}
  \begin{array}{rl}
    \avg{k^2}_{\rm DEG} & =
    \displaystyle{\sum_{k'=0}^{N-1}{k'}^2\widetilde{P}_{\rm
        DEG}(k')}\\ & =\displaystyle{
      \frac{1}{r}\sum_{k=\widetilde{k}}^{N-1}P(k)\sum_{k'=0}^{rN-1}{k'}^2\binom{k}{k'}\widetilde{r}^{k'}(1-\widetilde{r})^{k-k'}}\\ &
    =
    \displaystyle{\frac{1}{r}\sum_{k=\widetilde{k}}^{N-1}P(k)\left[k\widetilde{r}
        + k(k-1) \widetilde{r}^2\right]} \\ & =
    \displaystyle{\frac{\widetilde{r}^2}{r}\left[(1-\widetilde{r})\avg{k}
        + \sum_{k=\widetilde{k}}^{N-1}k^2P(k)\right]}
  \end{array}
    \label{eq:k2_avg_degree}
\end{equation}

\subsection*{Reduction of $R_{0}$ under ideal CT. }

We derive here a general expression for the effective value of the
expected number of secondary infections caused by a single infection
in a graph with perfect contact tracing, under the assumption that a
fraction $r$ of the nodes has installed a CT app. Assuming that a
generic node $\ell$ is infected, we want to estimate what is the
number of secondary infections caused by a node $i$ infected by
$\ell$.  The number of neighbours $R_i$ that can be infected by $i$
depends on whether $i$ has the CT app installed, and on how many of
its neighbours have their app installed. In particular, if $i$ does
not have the app, $R_i = \frac{\beta}{\mu}(k_i - 1)$, since we have to
remove the neighbour from which $i$ got the disease. If $i$ has a CT
app installed, instead, there are two possible cases:
\begin{enumerate}
\item
  If the node $\ell$ who infected $i$ has the CT app, then the
  infection has been ``detected'' by the app and $i$ goes into
  self-isolation immediately. If this happens, $i$ will not produce
  any secondary infection in the graph
\item
  If $i$ got infected by a neighbour without CT app, then the
  infection remains undetected, and $i$ can potentially infect $R_i =
  \frac{\beta}{\mu}(k_i-1)$ more nodes.
\end{enumerate}
So in the end, the expected number of infections potentially caused by
the infection of $i$ is:
\begin{equation}
  R_{i} = \frac{\beta}{\mu}(k_i-1) \frac{k_i-k_i'}{k_i}
\end{equation}
where the term $\frac{k_i-k_i'}{k_i}$ is the probability that the
infection of node $i$ does not get detected by the CT system. Hence,
the expected number of secondary infections potentially caused by
$\ell$ is given by:
\begin{equation}
  R_{\ell} = \frac{\beta}{\mu}\frac{1}{k_{\ell}}\sum_{i}a_{\ell i}(k_i-1) \frac{k_i-k_i'}{k_i}
\end{equation}
where $a_{\ell i}$ are the entries of the adjacency matrix of $G$. By
averaging $R_{\ell}$ over all the nodes of the graph we get the value
of the basic reproduction number with contact tracing:
\begin{equation}
  \begin{array}{rl}
    R'_{0} & =\displaystyle{
      \frac{\beta}{\mu N}\sum_{\ell}\frac{1}{k_{\ell}}\sum_{i}a_{\ell,i}R_i}\\ &
    =
    \displaystyle{\frac{\beta}{\mu N}\sum_{\ell}\frac{1}{k_{\ell}}\sum_{i}a_{\ell
        i}(k_i-1) \frac{k_i-k_i'}{k_i}} \\ & =
    \displaystyle{\frac{\beta}{\mu N}\left[\sum_{\ell}\frac{1}{k_{\ell}}\sum_{i}a_{\ell
          i} (k_i - 1) - \right.}\\
 &     \displaystyle{\left.  \sum_{\ell}\frac{1}{k_{\ell}}\sum_{i}a_{\ell
            i} \frac{k'_i}{k_i}(k_i - 1)\right]}
  \end{array}
  \label{eq:R_0_theo_method}
\end{equation}
This equation holds in general for any graph with any CT strategy.
Notice that the quantity:
\begin{equation}
  \frac{1}{N}\sum_{\ell}\frac{1}{k_{\ell}}\sum_{i}a_{\ell i} (k_i - 1)
\end{equation}
is the expected excess degree of the neighbours of a randomly sampled
node of $G$. In other words, it is equal to $\avg{k_{nn}(i)} - 1$,
where $k_{nn}(i)$ is the average degree of the neighbours of node
$i$. The basic reproduction number of the original graph $G$ is equal
to
\begin{equation}
  R_0 =
  \frac{\beta}{\mu}\frac{1}{N}\sum_{\ell}\frac{1}{k_{\ell}}\sum_{i}a_{\ell
      i} (k_i - 1)
\end{equation}
that is, the average degree of the neighbours of a randomly selected
nodes of $G$, multiplied by $\frac{\beta}{\mu}$. Hence we can
conveniently rewrite Eq.~(\ref{eq:R_0_theo_method}) as:
\begin{equation}
  R'_{0} = R_{0} - \frac{1}{N}\frac{\beta}{\mu}
  \sum_{\ell}\frac{1}{k_{\ell}}\sum_{i}a_{\ell i} \frac{k'_i}{k_i}(k_i
  - 1)
  \label{eq:recipe}
\end{equation}
In the special case when $G$ has no degree-degree correlations, we
have:
\begin{equation}
  \avg{k_{nn}(i)} =\frac{\avg{k^2}}{\avg{k}}\quad\quad \forall i
\end{equation}
hence we can write:
\begin{equation}
  R_0 \stackrel{\rm nc}{=}\frac{\beta}{\mu}\left[ \frac{\avg{k^2}}{\avg{k}} - 1\right]
\end{equation}
and we can rewrite Eq.~(\ref{eq:R_0_theo_method}) as:
\begin{equation}
  R'_{0} \stackrel{\rm nc}{=}
  \frac{\beta}{\mu}\left[\frac{\avg{k^2}}{\avg{k}} - 1 - \frac{1}{N}
      \sum_{\ell}\frac{1}{k_{\ell}}\sum_{i}a_{\ell i}
      \frac{k'_i}{k_i}(k_i - 1)\right]
  \label{eq:R_0_theo_nc}
\end{equation}
As expected, the effect of contact tracing is to reduce the basic
reproduction number of the original graph. In general,
Eq.~(\ref{eq:recipe}) (or Eq.~(\ref{eq:R_0_theo_nc}) in uncorrelated
graphs) provides a recipe to maximise the impact of CT app
installation. Indeed, given a certain adoption rate $r$, we can use
any optimisation algorithm to maximise the fitness function
\begin{equation}
  \max_{G'} \quad \mathcal{F}(G') =
  \sum_{\ell}\frac{1}{k_{\ell}}\sum_{i}a_{\ell i} \frac{k'_i}{k_i}(k_i
  - 1)
\end{equation}
over the ensemble of of the possible choices of $G'$. Notice that
$\mathcal{F}(G')$ can be decomposed in two terms. The first one is
\begin{equation}
  \sum_{\ell}\frac{1}{k_{\ell}}\sum_{i}a_{\ell i}k'_i
\end{equation}
that is, the sum of the expected degrees in the induced subgraph $G'$
of the neighbours of nodes in $G$, while the second one is:
\begin{equation}
  \sum_{\ell}\frac{1}{k_{\ell}}\sum_{i}a_{\ell i} \frac{k'_i}{k_i}
\end{equation}
i.e., the sum of the average fraction of degree in $G'$ and degree in
$G$ of all the neighbours of nodes in $G$. In the following we denote
the fraction of degree in $G'$ and degree in $G$ for node $i$ by $c_i
= \frac{k'_i}{k_i}$

\subsection*{Data description and set of networks studied}

In this work we considered 84 unique contact network data sets
constructed from two different types of data: (i) temporal network
data, which provide information regarding the different contacts
between individuals and the duration of each interaction; (ii) static
network data, where the contacts have been already aggregated for the
whole duration and a corresponding weight is associated to each
link. We reconstructed each network considering two distinct filtering
thresholds by either time -- in seconds for type (i) -- or by fraction
of links retained -- weight values for type (ii) --, resulting in 168
unique graphs.

For the networks of type (i), we considered a
hospital~\cite{vanhems2013} and a high school~\cite{salathe2010high}
temporal data sets, from which we filtered the contacts by applying
thresholds of 240 and 360 seconds, i.e., each temporal snapshots
resulted in a distinct network.  For an art gallery~\cite{isella2011s}
we used the thresholds of 0 and 20 seconds. Notice that we selected
these threshold values since they provide the largest connected
component for each network.  The type (ii) networks obtained from the
``Sociopatterns'' project include the contacts between individuals
with a weight that corresponds to either the number of contacts or
their duration~\cite{Sociopatterns,Genois2015,Genois2018}.  Given that
most type (ii) networks are densely connected and a significant
proportion of the weights have small values, we filter the networks by
keeping the top $25\%$ and $10\%$ links with the largest weights. A
list of all networks considered here is reported in Supplementary
Material Table 1.

\section*{Acknowledgements}

A.B. and V.N. acknowledge support from EPSRC New Investigator Award
grant no. EP/S027920/1. A.S. and V.N. acknowledge support from the
EPSRC Impact Acceleration Award -- Large Award Competition
programme. This work made use of the MidPLUS cluster, EPSRC grant
no. EP/K000128/1. This research used Queen Mary’s Apocrita HPC
facility, supported by QMUL Research-IT.
doi.org/10.5281/zenodo.438045.

\section*{Author contributions}
All the authors conceived the study. A.B. and A.S. performed the
numerical simulations, analysed the results, and prepared the figures.
S.S. and S.R. contributed to the methods for the analysis of the
results and prepared the figures. V.N. provided methods for the
analysis of the results and performed the analytical derivations. All
the authors wrote the manuscript and approved it in its final form.

\section*{Competing interests}

The authors declare no competing interests.

%\bibliographystyle{naturemag}
%\bibliography{epidemic_bibliography.bib}

%%Supplementary Material

\clearpage

\renewcommand{\figurename}{{\normalsize Supplementary Figure}}
\renewcommand{\tablename}{{\normalsize Supplementary Table}}
\renewcommand{\thefigure}{\arabic{figure}}
\renewcommand{\thesection}{\large Supplementary Note \arabic{section}}

\renewcommand{\thetable}{\arabic{table}}
\renewcommand{\theequation}{\arabic{equation}}
\renewcommand{\thepage}{\arabic{page}}

\setcounter{section}{0}
\setcounter{table}{0}
\setcounter{figure}{0}
\setcounter{equation}{0}

\onecolumngrid

\begin{center}
    \LARGE
    Supplementary Information for\\``Optimising the mitigation of epidemic spreading through targeted
        adoption of contact tracing apps''
    
    \normalsize
  
\end{center}

\maketitle

\section{Properties of the subgraphs induced by app installation.}

We provide here the full derivation of the degree distribution of the
subgraph $G'$ induced by a specific set of app installations, i.e.,
the subgraph of $G$ obtained by considering only the nodes which have
installed the app and the edges among them. We consider both the case
of uniform random and degree-based installation.

\subsection*{Uniform random installation}

We start by considering the case of uniform random installation on a
graph $G$ with assigned degree distribution $P(k)$, i.e., when a
fraction $r$ of the $N$ nodes, chosen uniformly at random, installs
the app. The probability that a given node $i$ that has installed the
app has degree $k'$ in $G'$ given the fact that it has degree $k$ in
the original graph $G$ can be expressed as:
\begin{equation}
P(k'_i = k'|k_i = k) = \binom{k}{k'} r^{k'}(1-r)^{k-k'}.
\label{eq:prob_condit_rand}
\end{equation}
Indeed, since app installation is performed uniformly at random with
probability $r$, the probability that a specific neighbour $j$ of a
node $i$ has installed the app is just $r$, and does not depend on
whether $i$ has installed the app or not. Consequently, the
probability that exactly $k'$ of the neighbours of $i$ have installed
the app is given by a Binomal distribution with success probability
$r$. This also implies that the expected degree of node $i$ in $G'$
across the ensemble of realisations of random installation is:
\begin{equation}
E[k'_i] = rk_i
\end{equation}
By using Eq.~(\ref{eq:prob_condit_rand}), we can express the
probability distribution $\widetilde{P}_{\rm RND}(k')$ that a generic
node with app installed has $k'$ neighbours which have installed the
app as:
\begin{equation}
\widetilde{P}_{\rm RND}(k') = \sum_{k=0}^{N-1}P(k)\binom{k}{k'}
r^{k'}(1-r)^{k-k'}
\label{eq:Pk_inst_rnd}
\end{equation}
Notice that this equation holds independently of the specific degree
distribution of $G$ and of other local properties of the graph, such
as the presence of degree-degree correlations or clustering. We can
derive the value of $\avg{k}_{\rm RND}$ by using the definition:
\begin{equation}
\begin{array}{rl}
\avg{k}_{\rm RND} & =
\displaystyle{\sum_{k=0}^{N-1}k\widetilde{P}_{\rm RND}(k) =
    \sum_{k=0}^{N-1}k
    \sum_{j=k}^{N-1}P(j)\binom{j}{k}r^{k}(1-r)^{j-k}} =
\displaystyle{\sum_{j=0}^{N-1}P(j)\sum_{k=0}^{N-1}k\binom{j}{k}r^{k}(1-r)^{j-k}}\\
& =
\displaystyle{\sum_{j=0}^{N-1}P(j)\sum_{k=0}^{j}k\binom{j}{k}r^{k}(1-r)^{j-k}}
= \displaystyle{\sum_{j=0}^{N-1}P(j)rj} \\
& \\
& = r\avg{k}_{G}
\end{array}
\label{eq:k_inst_rnd}
\end{equation}
Similarly, for $\avg{k^2}_{\rm RND}$ we obtain:
\begin{equation}
\begin{array}{rl}
\avg{k^2}_{\rm RND} & =
\displaystyle{\sum_{k=0}^{N-1}k^2\widetilde{P}_{\rm RND}(k) =
    \sum_{k=0}^{N-1}k^2
    \sum_{j=k}^{N-1}P(j)\binom{j}{k}r^{k}(1-r)^{j-k}} =
\displaystyle{\sum_{j=0}^{N-1}P(j)\sum_{k=0}^{N-1}k^2\binom{j}{k}r^{k}(1-r)^{j-k}}\\ &
=
\displaystyle{\sum_{j=0}^{N-1}P(j)\sum_{k=0}^{j}k^2\binom{j}{k}r^{k}(1-r)^{j-k}}
= \displaystyle{\sum_{j=0}^{N-1}P(j)\left[j(j-1)r^2 +
    rj\right]}\\ & \\ & = r^2\avg{k^2}_{G} + r(1-r)\avg{k}_{G}
\end{array}
\label{eq:k2_inst_rnd}
\end{equation}

\subsection*{Degree-based strategy and rich-club coefficient}

For degree-targeted app installations, i.e., when the top $rN$ nodes
in the ranking by degree have their app installed,
Eq.~(\ref{eq:prob_condit_rand}) does not hold, since the probability
of a node being in $G'$ is not uniform and depends instead on its
degree in $G$. If we call $\widetilde{k}$ the smallest of the degrees
of the nodes in $G'$, the subgraph induced by app installation
corresponds to the subgraph among nodes whose degree is $\ge
\widetilde{k}$. The fraction of existing edges among nodes with degree
$\ge \widetilde{k}$ is, by definition, the (unnormalised) rich-club
coefficient~\cite{colizza2006richclub_SI} of $G$:
\begin{equation}
\phi(k) = \frac{2e_{\ge k,\ge k}}{N_{\ge k}(N_{\ge k} - 1)}
\end{equation}
computed for $k = \widetilde{k}$. Here we denote by $N_{\ge k}$ the
number of nodes whose degree is larger than or equal to $k$, and by
$e_{\ge k,\ge k}$ the number of edges among those nodes. Consequently,
the average degree of $G'$ can be written as:
\begin{equation}
\avg{k}_{\rm DEG} = (N_{\ge\widetilde{k}} - 1) \phi(\widetilde{k})
\label{eq:avgk_g_prime}
\end{equation}

The general expression for the rich club in networks depends only on
the joint degree-degree distribution $P(k_1, k_2)$, i.e., the
probability of finding an edge between two nodes having degree $k_1$
and $k_2$, and reads~\cite{latora2017_SI,newman2018_SI}:
\begin{equation}
\phi(k) =
\frac{N\avg{k}\sum_{k_1,k_2=k}^{N-1}P(k_1,k_2)}{\left[N\sum_{k_1=k}^{N-1}P(k_1)\right]
    \left[\left(N\sum_{k_1=k}^{N-1}P(k_1)\right) - 1\right]}
\end{equation}
So in the case of degree-based targeting, even the first moment of the
degree distribution of $G'$ depends heavily on the presence of
degree-degree correlation in $G$, at stark difference with the case of
uniformly random installation seen above.

In the special case of uncorrelated random graphs, such as in the
configuration model ensemble, the joint degree-degree distribution
factorises as:
\begin{equation}
P(k_1,k_2) \stackrel{{\rm nc}}{=}
q_{k_1}q_{k_2} = \frac{k_1 k_2 P(k_1) P(k_2)}{\avg{k}^2}
\end{equation}
and it is easy to show that the rich club coefficient can be written
as:
\begin{equation}
\phi^{\rm nc}(k) =
\frac{\sum_{k_1=k}^{N-1}k_1P(k_1)\sum_{k_2=k}^{N-1}k_2P(k_2)}{\avg{k}
    \left[\sum_{k_1=k}^{N-1}P(k_1)\right]\left[N\sum_{k_2=k}^{N-1}P(k_2)-1\right]}
\end{equation}
By using Eq.~(\ref{eq:avgk_g_prime}) we obtain:
\begin{equation}
\avg{k}^{\rm nc}_{\rm DEG} = \left[N\sum_{k_1=\widetilde{k}}^{N-1}P(k_1) -
1\right]\phi^{\rm nc}(\widetilde{k}) =
\frac{\left[\sum_{k_1=\widetilde{k}}^{N-1}k_1P(k_1)\right]^2}{\avg{k}\left[\sum_{k_1=\widetilde{k}}^{N-1}P(k_1)\right]}
\label{eq:k_inst_deg}
\end{equation}
It is quite interesting to find that the first moment of the degree
distribution of $G'$ is indeed connected with the rich-club
coefficient of the graph at the critical degree.

It is actually possible to compute the full degree distribution of
$G'$ in the case of degree-based installation. If a node $i$ is in
$G'$, then we have $k_i \ge \widetilde{k}$. Now, the probability that
one of the $k_i$ neighbours of $i$ is in $G'$ is equal to:
\begin{equation}
Q_{\widetilde{k}}(i) = \sum_{k=\widetilde{k}}^{N-1} P(k|k_i)
\end{equation}
where $P(k|k_i)$ is the conditional probability of finding a node of
degree $k$ by following one of the edges of a node of degree $k_i$. In
the special case of graphs with no degree-degree correlations,
$P(k|k_i)$ does not depend on $k_i$, and factorises as:
\begin{equation}
P(k|k_i) \stackrel{\rm nc}{=} \frac{kP(k)}{\avg{k}}= q_k
\label{eq:no_corr_1}
\end{equation}
so we have:
\begin{equation}
Q_{\widetilde{k}}(i) \stackrel{\rm nc}{=} \sum_{k=\widetilde{k}}^{N-1}
q_k  =  \widetilde{r} \quad \forall i
\label{eq:no_corr_2}
\end{equation}
In the absence of degree-degree correlations, the probability of any
two specific nodes to be connected does not depend on their degree, by
definition. Hence, the probability that a node of $G'$ has a degree
equal to $k'$ is given again by the Binomial distribution:
\begin{equation}
\widetilde{P}(k'_i = k' | k_i = k) =
\binom{k}{k'} \widetilde{r}^{k'}(1- \widetilde{r})^{k-k'},
\quad k\geq \widetilde{k}
\end{equation}
while $\widetilde{P}(k'_i = k' | k_i = k)=0$ if
$k<\widetilde{k}$. Notice that $\widetilde{r}$ has the same role that
$r$ has in the equations for random assignment. In particular, this
means that the expected value $E\left[k_i'\right]$ across all the
configuration model graphs with a pre-assigned degree sequence is
equal to:
\begin{equation}
E\left[k_i'\right] = \widetilde{r}k_i
\label{eq:expected_k_prime}
\end{equation}
With an argument in all similar to that used for random installation,
we obtain:
\begin{equation}
\widetilde{P}(k') = \sum_{k=\widetilde{k}}^{N-1}P(k)
\binom{k}{k'}\widetilde{r}^{k'}(1-\widetilde{r})^{k-k'}
\label{eq:Pk_inst_degbased_SI}
\end{equation}

Notice that $\widetilde{P}(k')$ represents the probability to find a
node of $G$ which has degree $k'$ in the subgraph induced by app
installations. To obtain the actual degree distribution in the induced
subgraph, i.e., the probability that one of the nodes of $G'$ has
degree $k'$, we must rescale $\widetilde{P}(k')$ to the nodes in $G'$,
i.e., we consider the probability distribution:
\begin{equation}
\widetilde{P}_{\rm DEG}(k') = \frac{1}{r}\widetilde{P}(k') =
\frac{1}{r}\sum_{k=\widetilde{k}}^{N-1}P(k)
\binom{k}{k'}\widetilde{r}^{k'}(1-\widetilde{r})^{k-k'}
\label{eq:Pk_deg_norm}
\end{equation}
It is important to stress here that the expression for
$\widetilde{P}_{\rm DEG}(k')$ provided above is valid only in
uncorrelated graphs, due to the assumption we made in
Eq.~(\ref{eq:no_corr_1}) and Eq.~(\ref{eq:no_corr_2}).

\begin{figure*}[b!]
    \includegraphics[width=6.2in]{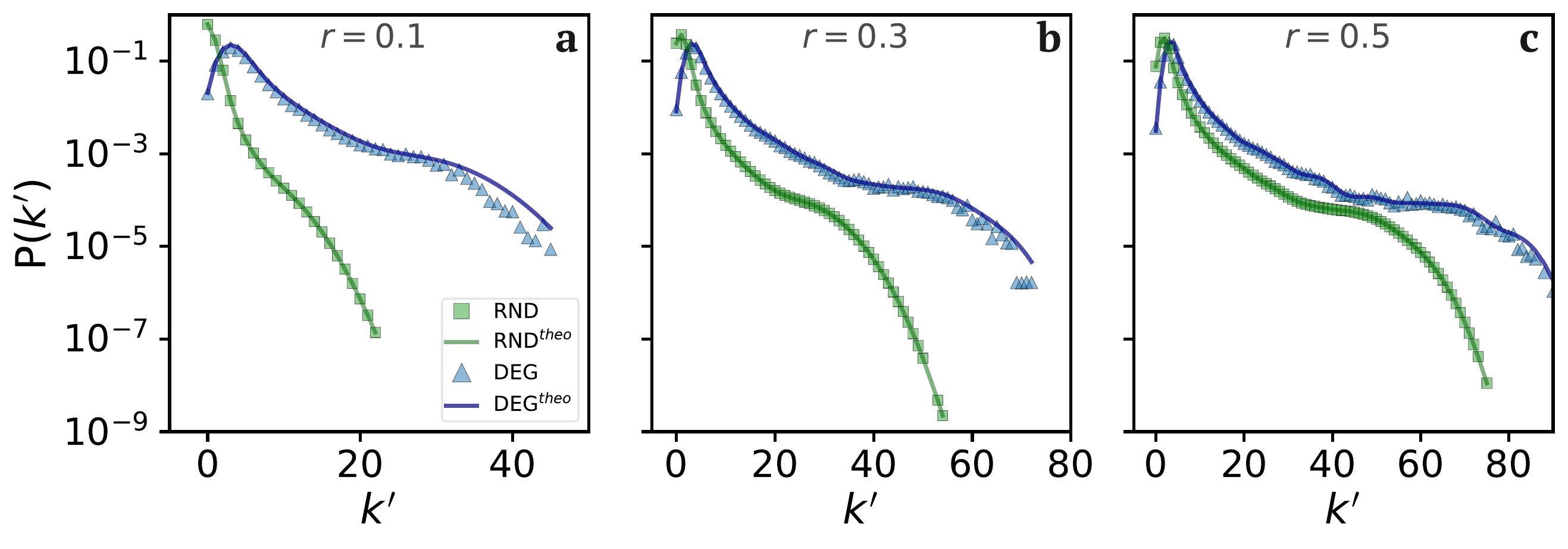}
    \caption[Empirical and theoretical degree distributions of the induced
    subgraph for different values of adoption rate r] {\textbf{Empirical
            and theoretical degree distributions of the induced subgraph for
            different values of adoption rate r. } We report the comparison
        between the empirical degree distributions of the induced subgraph
        $G'$ for the random (RND) and degree-based (DEG) strategies when
        considering three different adoption rates $r$. Solid lines
        represent the theoretical predictions for the RND and DEG strategies
        calculated using Eq.~(\ref{eq:Pk_inst_rnd}) and
        Eq.~(\ref{eq:Pk_deg_norm}), respectively. The plots correspond
        to an ensemble of configuration model graphs with degree
        distribution $P(k) \sim k^{-3}$ and $N =10^4$ nodes. Results
        averaged over $10^4$ realisations.}
    
    \label{fig:SI_fig1}
\end{figure*}

\medskip
In Supplementary Fig.~\ref{fig:SI_fig1} we report the the empirical degree
distributions of the subgraph $G'$ induced by the random (RND) and
degree-based (DEG) CT strategies when considering three different
adoption rates $r=0.1, 0.3, 0.5$, respectively. We considered here an
ensemble of configuration model graphs with degree distribution $P(k)
\sim k^{-3}$ and $N =10^4$ nodes. Notice that the numerical
simulations are in perfect agreement with the analytical predictions
in Eq.~(\ref{eq:Pk_inst_rnd}) and Eq.~(\ref{eq:Pk_deg_norm})

\subsection*{Simulated annealing}
The simulated annealing (SA) procedure is commonly used to find the global
optimum of a certain cost/energy function. In a SA algorithm, an energy
function usually attributes certain values to each configuration of the
system. The best configuration, which is the one that optimises the energy
function, is usually identified by searching the phase space of the system
considering Markov Chain Monte Carlo moves that allow to switch from one
configuration to another. In our context, we are interested on finding the
maximum of the following cost function:
\begin{equation}
\mathcal{F}(G') =
\sum_{\ell}\frac{1}{k_{\ell}}\sum_{i}a_{\ell i} \frac{k'_i}{k_i}(k_i
- 1)
\label{eq:cost_function_SI}
\end{equation}
Hence, for a given app adoption rate $r$, the SA algorithm tries to find
the nodes which provide the maximum value of $\mathcal{F}$ in order to
reduce the expected number of secondary infections caused by a single
contagion in presence of contact tracing. Each configuration is represented
by the couple $\{A,f\}$, where $A$ is set of nodes ID which adopts the CT
app in the network, while $f$ is the energy associated with the
configuration, i.e., the value computed using Eq.~\ref{eq:cost_function_SI}.

For a given app adoption rate $r$, we start at time $t=0$ from a random
configuration $A_0$ and evaluate the corresponding energy $f_0$. Then, at
each step $t$ of the algorithm, we randomly replace a node $i$ of the set
$A_{t-1}$ with a randomly selected node $j$ that does not belong to the
same set, i.e. $A_{t}=(A_{t-1} \cup \{j\}) \setminus \{i\}$. After
calculating the energy $f_t$ of the new configuration, we accept it
with probability:
\begin{equation*}
p =
\begin{cases}
1 & \text{ if } f_t > f \\
{\rm e}^{-\frac{f-f_t}{T}}
& \text{otherwise}
\end{cases}
\end{equation*}
where $T$ has the role of temperature in the simulated annealing procedure.
In particular, the initial temperature for the simulations was set to
$T_{max} = 1$ and in every step it was reduced by $\delta T$ until
$T_{min} = 0.01$ was reached. A step of $\delta T = 10^{-7}$ was used for
all our numerical simulations.

\cleardoublepage
\section{Contact tracing strategies based on local information}

We report in Supplementary Table \ref{Table1} the real-world networks
analysed in the main manuscript along with the thresholds values used
for filtering nodes or edges in the networks and basic statistics
(resulting number of nodes and edges). For each of the $84$ unique
graphs analysed, we report the thresholds used and the resulting number
of links with larger weights.

\begin{center}
    \begin{table*}[ht!]
        \renewcommand*{\arraystretch}{0.95}
        \resizebox{\textwidth}{!}{
                        \scriptsize
            \begin{tabular}{|l|l|l|l|l|l|l|l|l|l|l|l|}
                \hline
                Network & Thres. & Nodes & Edges & Network & Thres. & Nodes & Edges & Network & Thres. & Nodes & Edges \\
                \hline
                contacts-Hospital & 240 sec. & 67 & 291 & contacts-Hospital & 360 sec. & 63 & 228 & highschool-2011 & 240 sec. & 117 & 332\\\hline
                highschool-2011 & 360 sec. & 111 & 252 & highschool-2012 & 240 sec. & 171 & 496 & highschool-2012 & 360 sec. & 158 & 372\\\hline
                gall-2009-04-28 & 0 sec. & 190 & 703 & gall-2009-04-28 & 20 sec. & 61 & 127 & gall-2009-04-29 & 0 sec. & 198 & 736\\\hline
                gall-2009-04-29 & 20 sec. & 112 & 231 & gall-2009-04-30 & 0 sec. & 144 & 486 & gall-2009-04-30 & 20 sec. & 59 & 132\\\hline
                gall-2009-05-01 & 0 sec. & 201 & 558 & gall-2009-05-01 & 20 sec. & 37 & 60 & gall-2009-05-02 & 0 sec. & 213 & 966\\\hline
                gall-2009-05-02 & 20 sec. & 89 & 163 & gall-2009-05-03 & 0 sec. & 305 & 1847 & gall-2009-05-03 & 20 sec. & 211 & 513\\\hline
                gall-2009-05-05 & 0 sec. & 78 & 147 & gall-2009-05-05 & 20 sec. & 18 & 26 & gall-2009-05-06 & 0 sec. & 176 & 745\\\hline
                gall-2009-05-06 & 20 sec. & 37 & 65 & gall-2009-05-07 & 0 sec. & 194 & 801 & gall-2009-05-07 & 20 sec. & 72 & 204\\\hline
                gall-2009-05-09 & 0 sec. & 216 & 993 & gall-2009-05-09 & 20 sec. & 156 & 312 & gall-2009-05-10 & 0 sec. & 168 & 625\\\hline
                gall-2009-05-10 & 20 sec. & 138 & 242 & gall-2009-05-12 & 0 sec. & 56 & 114 & gall-2009-05-12 & 20 sec. & 11 & 15\\\hline
                gall-2009-05-13 & 0 sec. & 166 & 590 & gall-2009-05-13 & 20 sec. & 58 & 82 & gall-2009-05-14 & 0 sec. & 132 & 620\\\hline
                gall-2009-05-14 & 20 sec. & 54 & 134 & gall-2009-05-15 & 0 sec. & 241 & 1301 & gall-2009-05-15 & 20 sec. & 127 & 334\\\hline
                gall-2009-05-16 & 0 sec. & 241 & 1504 & gall-2009-05-16 & 20 sec. & 216 & 577 & gall-2009-05-17 & 0 sec. & 187 & 1347\\\hline
                gall-2009-05-17 & 20 sec. & 172 & 470 & gall-2009-05-19 & 0 sec. & 49 & 112 & gall-2009-05-19 & 20 sec. & 13 & 23\\\hline
                gall-2009-05-20 & 0 sec. & 89 & 507 & gall-2009-05-20 & 20 sec. & 75 & 276 & gall-2009-05-21 & 0 sec. & 43 & 193\\\hline
                gall-2009-05-21 & 20 sec. & 38 & 107 & gall-2009-05-22 & 0 sec. & 131 & 864 & gall-2009-05-22 & 20 sec. & 65 & 183\\\hline
                gall-2009-05-23 & 0 sec. & 238 & 1075 & gall-2009-05-23 & 20 sec. & 208 & 396 & gall-2009-05-24 & 0 sec. & 31 & 68\\\hline
                gall-2009-05-24 & 20 sec. & 10 & 13 & gall-2009-05-26 & 0 sec. & 131 & 513 & gall-2009-05-26 & 20 sec. & 57 & 183\\\hline
                gall-2009-05-27 & 0 sec. & 116 & 395 & gall-2009-05-27 & 20 sec. & 45 & 118 & gall-2009-05-28 & 0 sec. & 141 & 1054\\\hline
                gall-2009-05-28 & 20 sec. & 100 & 495 & gall-2009-05-29 & 0 sec. & 93 & 272 & gall-2009-05-29 & 20 sec. & 65 & 132\\\hline
                gall-2009-05-30 & 0 sec. & 127 & 397 & gall-2009-05-30 & 20 sec. & 70 & 121 & gall-2009-05-31 & 0 sec. & 89 & 267\\\hline
                gall-2009-05-31 & 20 sec. & 22 & 56 & gall-2009-06-02 & 0 sec. & 16 & 61 & gall-2009-06-02 & 20 sec. & 11 & 41\\\hline
                gall-2009-06-03 & 0 sec. & 62 & 174 & gall-2009-06-03 & 20 sec. & 16 & 46 & gall-2009-06-04 & 0 sec. & 53 & 382\\\hline
                gall-2009-06-04 & 20 sec. & 37 & 137 & gall-2009-06-05 & 0 sec. & 88 & 267 & gall-2009-06-05 & 20 sec. & 32 & 56\\\hline
                gall-2009-06-06 & 0 sec. & 142 & 696 & gall-2009-06-06 & 20 sec. & 127 & 324 & gall-2009-06-07 & 0 sec. & 155 & 563\\\hline
                gall-2009-06-07 & 20 sec. & 113 & 203 & gall-2009-06-09 & 0 sec. & 74 & 238 & gall-2009-06-09 & 20 sec. & 21 & 70\\\hline
                gall-2009-06-10 & 0 sec. & 35 & 74 & gall-2009-06-10 & 20 sec. & 18 & 49 & gall-2009-06-11 & 0 sec. & 77 & 161\\\hline
                gall-2009-06-11 & 20 sec. & 13 & 44 & gall-2009-06-12 & 0 sec. & 58 & 158 & gall-2009-06-12 & 20 sec. & 25 & 101\\\hline
                gall-2009-06-13 & 0 sec. & 102 & 264 & gall-2009-06-13 & 20 sec. & 21 & 40 & gall-2009-06-14 & 0 sec. & 138 & 433\\\hline
                gall-2009-06-14 & 20 sec. & 105 & 181 & gall-2009-06-16 & 0 sec. & 67 & 391 & gall-2009-06-16 & 20 sec. & 48 & 198\\\hline
                gall-2009-06-17 & 0 sec. & 72 & 212 & gall-2009-06-17 & 20 sec. & 48 & 81 & gall-2009-06-18 & 0 sec. & 74 & 275\\\hline
                gall-2009-06-18 & 20 sec. & 23 & 66 & gall-2009-06-19 & 0 sec. & 125 & 412 & gall-2009-06-19 & 20 sec. & 64 & 112\\\hline
                gall-2009-06-20 & 0 sec. & 149 & 495 & gall-2009-06-20 & 20 sec. & 109 & 203 & gall-2009-06-21 & 0 sec. & 166 & 676\\\hline
                gall-2009-06-21 & 20 sec. & 123 & 244 & gall-2009-06-23 & 0 sec. & 57 & 128 & gall-2009-06-23 & 20 sec. & 19 & 31\\\hline
                gall-2009-06-24 & 0 sec. & 79 & 369 & gall-2009-06-24 & 20 sec. & 22 & 40 & gall-2009-06-25 & 0 sec. & 79 & 321\\\hline
                gall-2009-06-25 & 20 sec. & 31 & 152 & gall-2009-06-26 & 0 sec. & 78 & 152 & gall-2009-06-26 & 20 sec. & 9 & 13\\\hline
                gall-2009-06-27 & 0 sec. & 35 & 99 & gall-2009-06-27 & 20 sec. & 10 & 17 & gall-2009-06-28 & 0 sec. & 107 & 397\\\hline
                gall-2009-06-28 & 20 sec. & 35 & 71 & gall-2009-06-30 & 0 sec. & 128 & 435 & gall-2009-06-30 & 20 sec. & 34 & 55\\\hline
                gall-2009-07-01 & 0 sec. & 167 & 814 & gall-2009-07-01 & 20 sec. & 127 & 336 & gall-2009-07-02 & 0 sec. & 60 & 180\\\hline
                gall-2009-07-02 & 20 sec. & 21 & 39 & gall-2009-07-03 & 0 sec. & 121 & 321 & gall-2009-07-03 & 20 sec. & 32 & 60\\\hline
                gall-2009-07-04 & 0 sec. & 127 & 526 & gall-2009-07-04 & 20 sec. & 103 & 214 & gall-2009-07-05 & 0 sec. & 95 & 314\\\hline
                gall-2009-07-05 & 20 sec. & 30 & 40 & gall-2009-07-07 & 0 sec. & 220 & 1187 & gall-2009-07-07 & 20 sec. & 166 & 477\\\hline
                gall-2009-07-08 & 0 sec. & 186 & 820 & gall-2009-07-08 & 20 sec. & 159 & 356 & gall-2009-07-09 & 0 sec. & 114 & 373\\\hline
                gall-2009-07-09 & 20 sec. & 61 & 121 & gall-2009-07-10 & 0 sec. & 157 & 776 & gall-2009-07-10 & 20 sec. & 103 & 359\\\hline
                gall-2009-07-11 & 0 sec. & 161 & 673 & gall-2009-07-11 & 20 sec. & 102 & 188 & gall-2009-07-12 & 0 sec. & 148 & 580\\\hline
                gall-2009-07-12 & 20 sec. & 114 & 215 & gall-2009-07-14 & 0 sec. & 275 & 1633 & gall-2009-07-14 & 20 sec. & 195 & 566\\\hline
                gall-2009-07-15 & 0 sec. & 410 & 2765 & gall-2009-07-15 & 20 sec. & 351 & 1205 & gall-2009-07-16 & 0 sec. & 318 & 1441\\\hline
                gall-2009-07-16 & 20 sec. & 250 & 567 & gall-2009-07-17 & 0 sec. & 221 & 1073 & gall-2009-07-17 & 20 sec. & 180 & 405\\\hline
                InVS13 & $25\%$ & 93 & 967 & InVS13 & $10\%$ & 87 & 391 & InVS15 & $25\%$ & 217 & 4150\\\hline
                InVS15 & $10\%$ & 208 & 1656 & LH10 & $25\%$ & 62 & 339 & LH10 & $10\%$ & 42 & 116\\\hline
                LyonSchool & $25\%$ & 240 & 6637 & LyonSchool & $10\%$ & 225 & 2654 & SFHH & $25\%$ & 392 & 17709\\\hline
                SFHH & $10\%$ & 384 & 7048 & Thiers13 & $25\%$ & 326 & 10873 & Thiers13 & $10\%$ & 69 & 1223\\\hline
                bt & $25\%$ & 634 & 18727 & bt & $10\%$ & 617 & 7449 & call & $25\%$ & 12 & 11\\\hline
                call & $10\%$ & 5 & 5 & enterprise & $25\%$ & 84 & 183 & enterprise & $10\%$ & 50 & 66\\\hline
                Hospital & $25\%$ & 67 & 284 & Hospital & $10\%$ & 44 & 110 & sms & $25\%$ & 19 & 18\\\hline
                sms & $10\%$ & 6 & 5 & student & $25\%$ & 1452 & 17134 & student & $10\%$ & 956 & 6670\\\hline
        \end{tabular}}
        \caption[Networks analysed and main statistics]
        {Networks analysed and main statistics.}\label{Table1}
    \end{table*}
\end{center}

Some of the strategies studied in the main manuscript, such as
degree-based targeting, require global and accurate information about
the contact network. Unfortunately, this requirement can represent a
major challenge in a realistic scenario, as one normally has access
only to partial or local information about a node and/or about its
immediate neighbours. To mitigate this problem, we explored two
decentralised strategies inspired by the friendship paradox in social
networks~\cite{feld1991your_SI,eom2014generalized_SI} and previously
considered for efficient
vaccination~\cite{cohen2003efficient_SI,lelarge2009efficient_SI,christakis2010social_SI}.
The paradox is that, on average, in any graph each individual is more
likely to have fewer friends than its own friends do, which
corresponds to the saying that ``your friends are more social than you
are''. We leverage this concept to construct node rankings based on a
simple ``voting'' system. In particular, we assume that each node $i$
in the network has the possibility to indicate one of its neighbours
as a candidate for app installation. Then, we rank each node $i$
according to the total number of votes $v_i$ it received.

Although the voting system requires each node to have access to just
local information, it is easy to show that if the graph has no
degree-degree correlations, then the number of votes received by a
node $i$ is actually proportional to its degree $k_i$. In particular
if the probability that a node $j$ casts a vote to a neighbour $i$ is
$1/k_j$, the number of votes received by a node $i$ is given by:
\begin{equation}
v_i=k_i\sum_{j} A_{ji}\frac{1}{k_j},
\label{eq:vote_theo}
\end{equation}
where $a_{ji}$ are the entried of the adjacency matrix of the graph.
Despite the number of votes received is proportional to the degree of
a node, the inflow also plays a very significant role -- similarly to
what happens in PageRank -- which can lead to some nodes with higher
degree actually receiving a smaller number of votes than nodes with a
lower degree. In Supplementary Fig. \ref{fig:SI_votes} we compare the
average number of votes received by each of the nodes in an ensemble
of configuration model graphs with degree distribution $P(k)\sim
k^{-3}$ and $N =10^4$ nodes and the theoretical prediction provide by
Eq.~(\ref{eq:vote_theo}). The agreement between the simulations and
the prediction is perfect, and confirms that indeed the number of
votes received by a node is somehow proportional to its degree.
\begin{figure*}[!htbp]
    \begin{center}
        \includegraphics[width=4.2in]{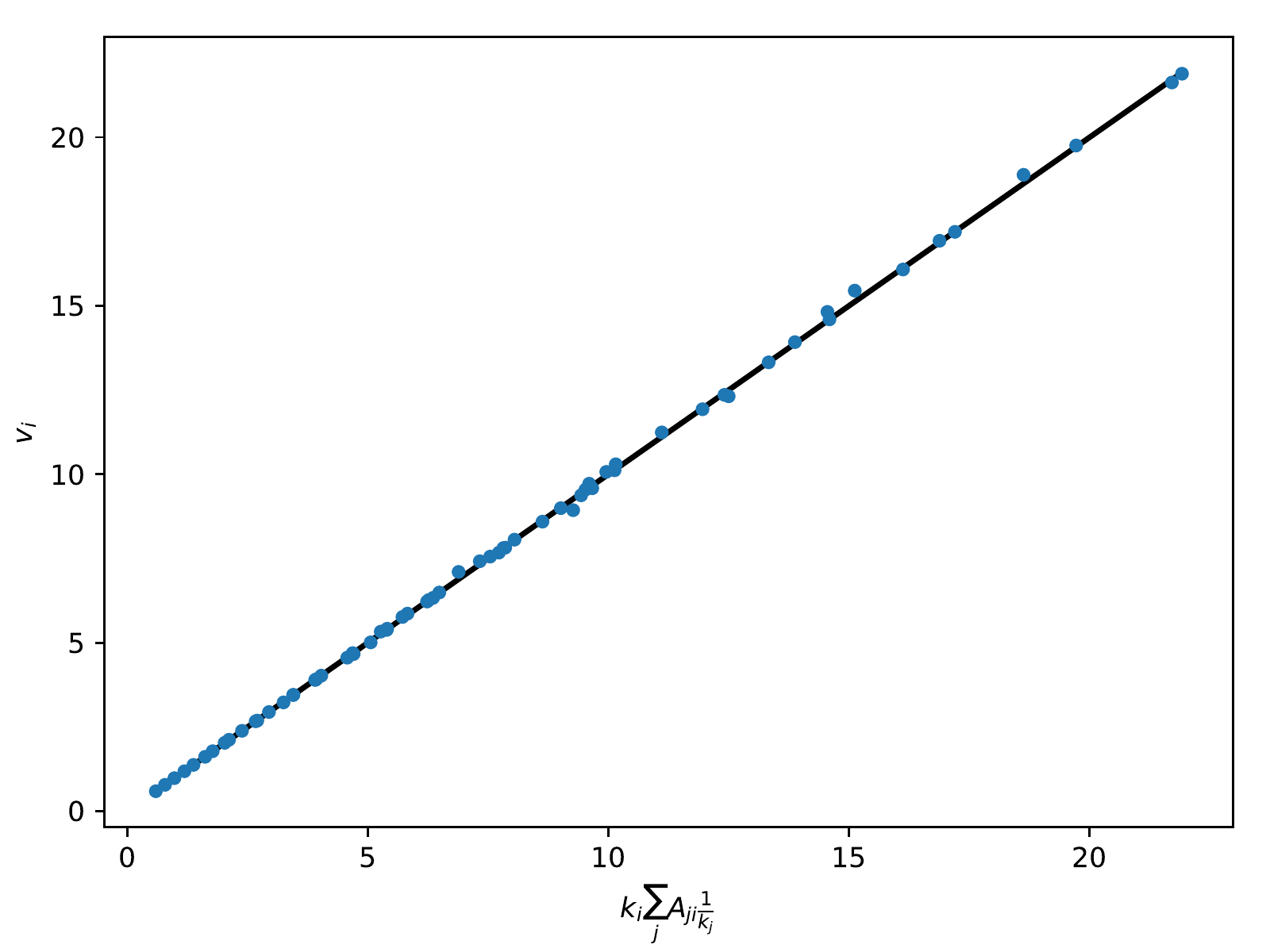}
    \end{center}
    \caption[Comparison between the votes received by a node and the
    theoretical prediction]{\textbf{Comparison between the votes received
            by a node and the theoretical prediction when each node casts a single
            vote.} We considered an ensemble of configuration model graphs with
        degree distribution $P(k)\sim k^{-3}$ and $N =10^4$ nodes. The votes
        received, and the theoretical prediction, have been averaged over all
        the nodes with the same degree and the black line indicates a perfect
        match between both.}
    \label{fig:SI_votes}
\end{figure*}

The assumption that individuals will install the CT app if they are
directly targeted, and with probability equal to 1, is rather simplistic
and difficult to achieve in real-world systems. Thus, we also explored
how effective the mitigation of an epidemic would be when instead of
following a strict adoption based on rankings, a node $i$ has a certain
probability $\sigma_i$ to adopt the app, which depends on the number of
votes it has received. One of the simplest approaches to model the adoption
of technologies is based on the usage of sigmoid adoption
functions~\cite{stoneman1981intra_SI,button1993modelling_SI,tsur1990dynamic_SI}
in the form:
\begin{equation}
\sigma_i(v_i,a)=\frac{1}{1+\exp(-a(v_i-\langle v \rangle))}
\end{equation}
where $\langle v\rangle$ is the average number of votes received by
nodes in a given realisation, $v_i$ is the number of votes received by
node $i$, and $a$ tunes the intensity of the sigmoid. In our
simulations we considered either $a=1$ or $a=5$.

It is important to note that the function $\sigma_i(v,a)$ needs to be
properly normalised by $\sum_i \sigma_i(v_i,a)$ to obtain a
probability function. In Supplementary Fig.~\ref{fig:SI_fig2}(a-c) and (d-f), we
report the temporal evolution of the number of infected individuals
$I(t)$ for the recommendation-based strategy, respectively for the
high school and the workplace network analysed in the main manuscript.

\begin{figure*}[!t]
    \begin{center}
        \includegraphics[width=6.2in]{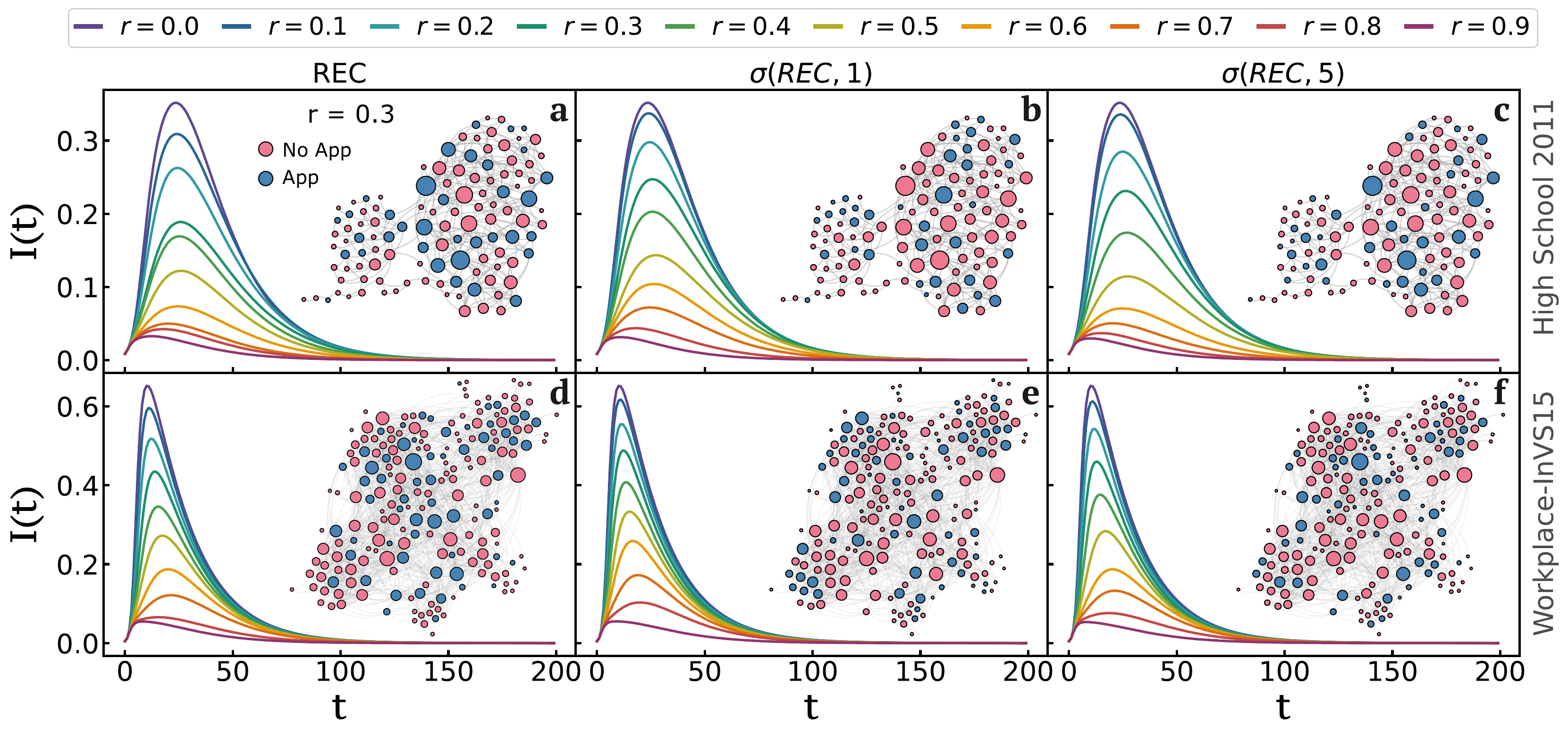}
    \end{center}
    \caption[Impact of CT strategy and adoption rate on the
    epidemic trajectory in real-world networks]{\textbf{Impact of CT strategy and adoption rate on the
            epidemic trajectory in real-world networks}. The evolution of
        the disease in a SIR model with contact tracing depends heavily
        on the adoption rate and on the strategy considered to select
        the nodes with CT apps. Here, we show the results on two
        real-world systems, namely, the high-resolution face-to-face
        contact data recorded for a high school (a-c) and a workplace
        network (d-f). The strategies analysed are
        the recommendation-based (a,d), the sigmoids $\sigma_i(v_i,a=1)$ (b,e),
        and $\sigma_i(v_i,a=5)$ (c,f).
        For these simulations we set $\beta=0.1$ and $\mu=0.05$.}
    \label{fig:SI_fig2}
\end{figure*}

\begin{figure*}[!ht]
    \begin{center}
        \includegraphics[width=6in]{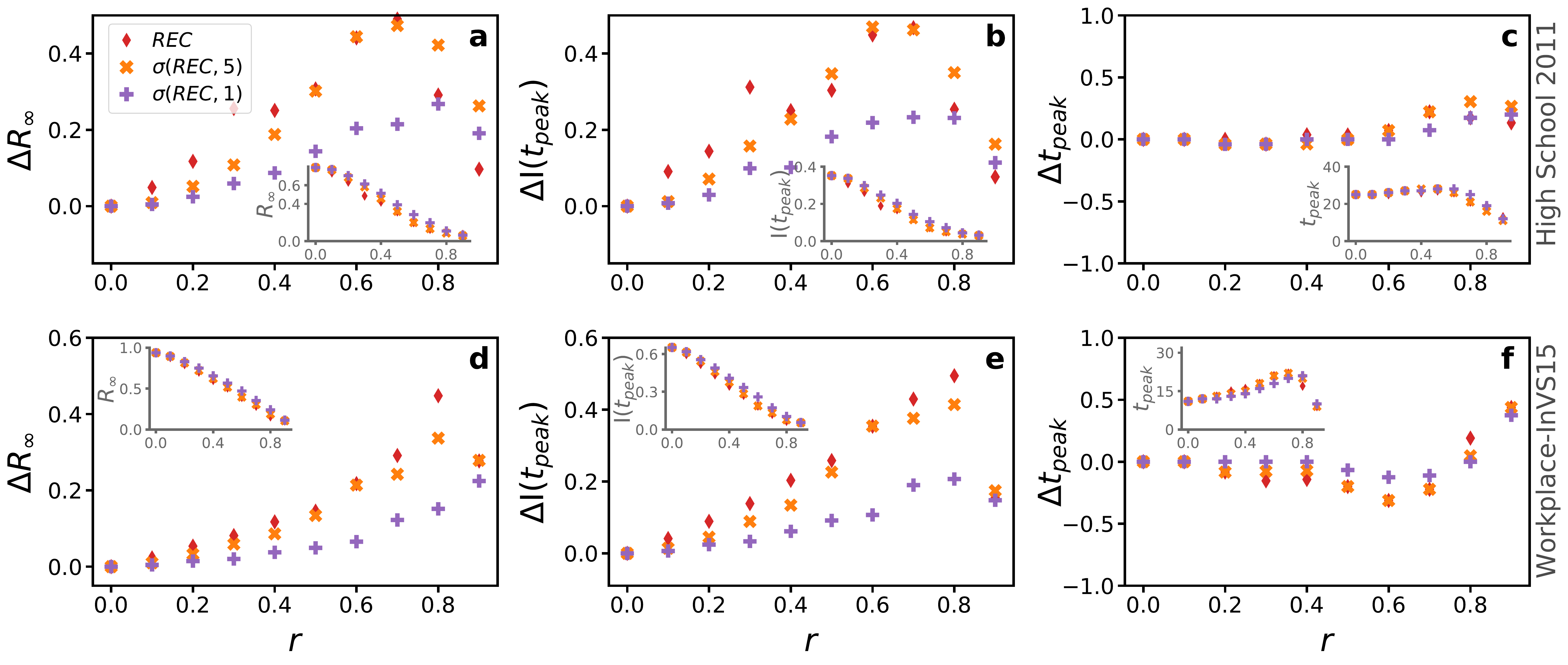}
    \end{center}
    \caption[Comparison of epidemic indicators under different
    CT strategies]{\textbf{Comparison of epidemic indicators under different
            CT strategies}. Relative decrease with respect to random
        installations of the total number of recovered $R_{\infty}$,
        height of the infection peak $I(t_{\rm peak})$ and position of the
        peak $t_{\rm peak}$ for the recommendation-based, the sigmoids $\sigma_i(v_i,a=1)$,
        and $\sigma_i(v_i,a=5)$ installation strategies in the high school (a-c) and
        workplace networks (d-f). The inset of
        each panel reports the plot of the raw variable, respectively
        $R_{\infty}$ (panel a and d), $I(t_{\rm peak})$ (panel b and e)
        and $t_{\rm peak}$ (panel c and f).}
    \label{fig:SI_fig3}
\end{figure*}

In the high school network we observe a more drastic reduction of
cases and larger difference between the strategies as the
level of app adoption increases. Furthermore, even though the
mitigation effect with the sigmoids is less pronounced than the best
case scenarios relying on global information, we still observe a
drastic reduction when compared to the random case. In particular, to
obtain a strong decrease in the total number of infected individuals
we need to consider an adoption rate $r$ above $0.5$. To complement
the previous results, Fig.~\ref{fig:SI_fig3} displays the relative
improvement over RND for the recommendation-based strategy and the
two sigmoids as a function of $r$ for the same two networks. In
particular, we focus on the total number $R_{\infty}$ of people who
got the disease, the maximum number $I(t_{\rm peak})$ of infected
across the duration of the epidemic, and the time $t_{\rm peak}$ at
which that number is achieved. Remarkably, for the high school network
there are no strong differences in performance between
$\sigma_i(v_i,a=5)$ and an assignment based on the ranking of votes
received. Finally, the position of the peak does not seem heavily
affected by the sigmoid strategies, so that there is still a visible
shift of the epidemic peak towards the left for high values of
$r$ in the high school network.

\begin{figure*}[!hb]
    \begin{center}
        \includegraphics[width=5.9in]{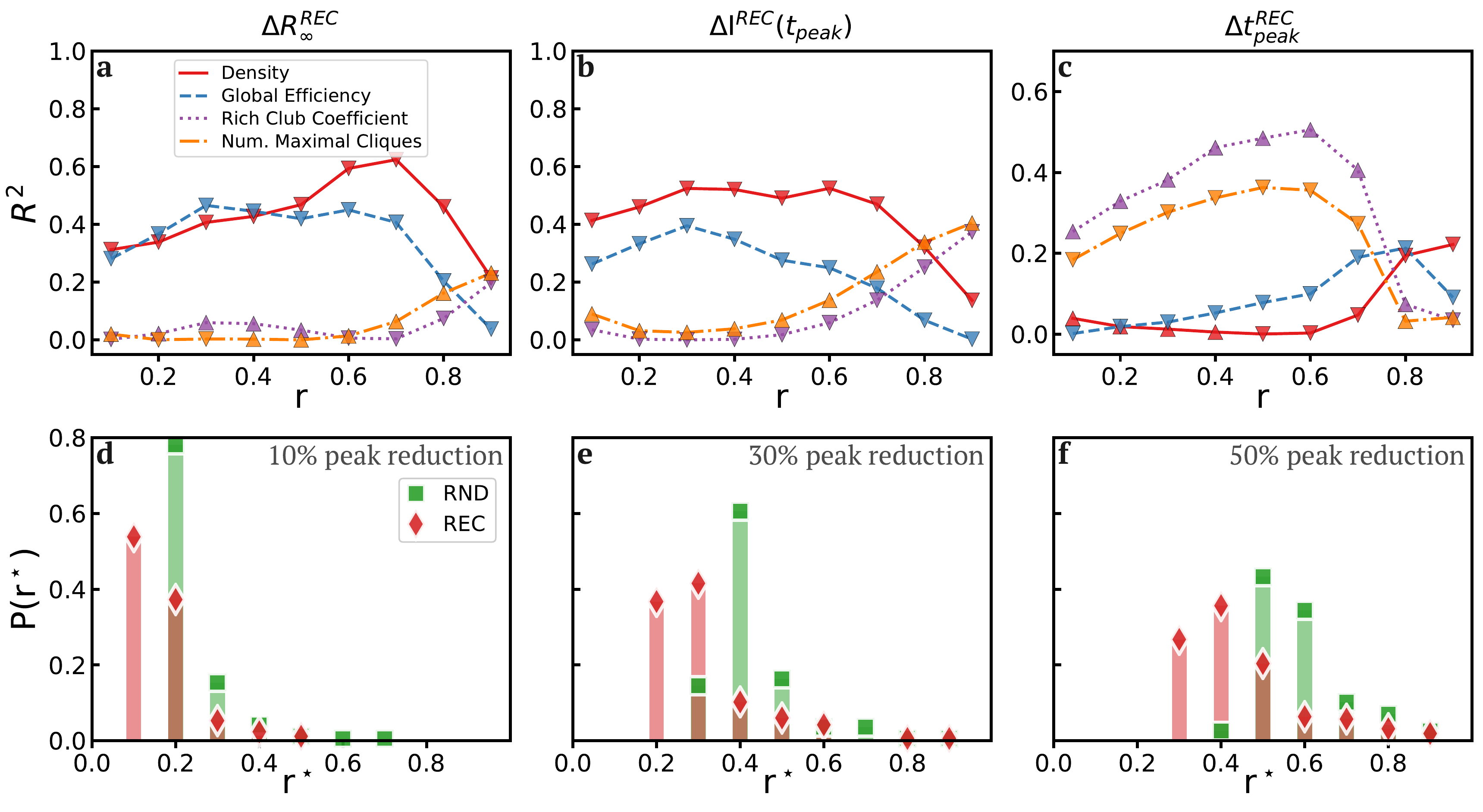}
    \end{center}
    \caption[Correlations with network structural measures
    and performance of CT strategies to mitigate an
    epidemic]{\textbf{Correlations with network structural measures and
            performance of CT strategies to mitigate an epidemic}. For the 168
        real-world contact networks analysed, panels (a-c) report the correlation
        ($R^2$) between the epidemiological indicators $\Delta R^{REC}_{\infty}$
        (a), $\Delta I^{REC}(t_{\rm peak})$ (b) and $\Delta t^{REC}_{\rm peak}$
        (c) and the network density (red), the global efficiency (blue), the rich
        club coefficient (purple) and the number of maximal cliques (orange) for
        different values of $r$. All measures were computed considering the full
        graphs. In panels (d-f), we report the distribution of minimum adoption
        ratios $r^{*}$ needed to produce a $10\%$ (d), $30\%$ (e) and $50\%$
        (f) for the REC (red) and RND (green) strategies. Overall, the REC
        strategy performs better than RND and requires lower adoption ratios
        to produce an equivalent reduction of the peak.}
    \label{fig:SI_fig4}
\end{figure*}

In the following, we refer to the local strategy based on the number of
votes $v$ received by a node at the recommendation-based strategy (REC).
Considering the 168 networks, Supplementary Fig.~\ref{fig:SI_fig4}(a-c)
reports the results with the correlations between the epidemiological
indicators $\Delta R^{REC}_{\infty}$, $\Delta I^{REC}(t_{\rm peak})$ and
$\Delta t^{REC}_{\rm peak}$ and structural measures as a function of the
adoption rate $r$. Overall, for low/moderate values of $r$, the performance
of the REC strategy -- measured by $\Delta R^{REC}_{\infty}$ and
$\Delta I^{REC}(t_{\rm peak})$ -- displays a moderate correlation only
with the density and global efficiency of the contact networks. Conversely,
the rich club coefficient and number of maximal cliques show a higher
correlation only when considering $\Delta t^{REC}_{\rm peak}$. As done
for the DTI, we have also compared how the REC strategy is able
to mitigate $I(t_{\rm peak})$ when compared to RND, as can be seen
in Supplementary Fig.~\ref{fig:SI_fig4}(d-f) where we report the
distribution of $r^{*}$ needed for a reduction of of 10\%, 30\% and
50\%, respectively.

Finally, to compare all the different decentralised CT strategies presented
in this work, we report in Supplementary Fig.~\ref{fig:SI_fig5} the epidemic
trajectory for the two real-world systems analysed in the main manuscript
when considering different CT adoption rates $r$. Overall, $DTI$ is the
best performing decentralised strategy, which provides a significant
reduction with respect to the random case both in terms of the evolution
of the epidemic spreading and for the height of the peak even for low
values of installation rate $r$. Such effect is notably visible in the
high school network.

\begin{figure*}[!h]
    \begin{center}
        \includegraphics[width=6in]{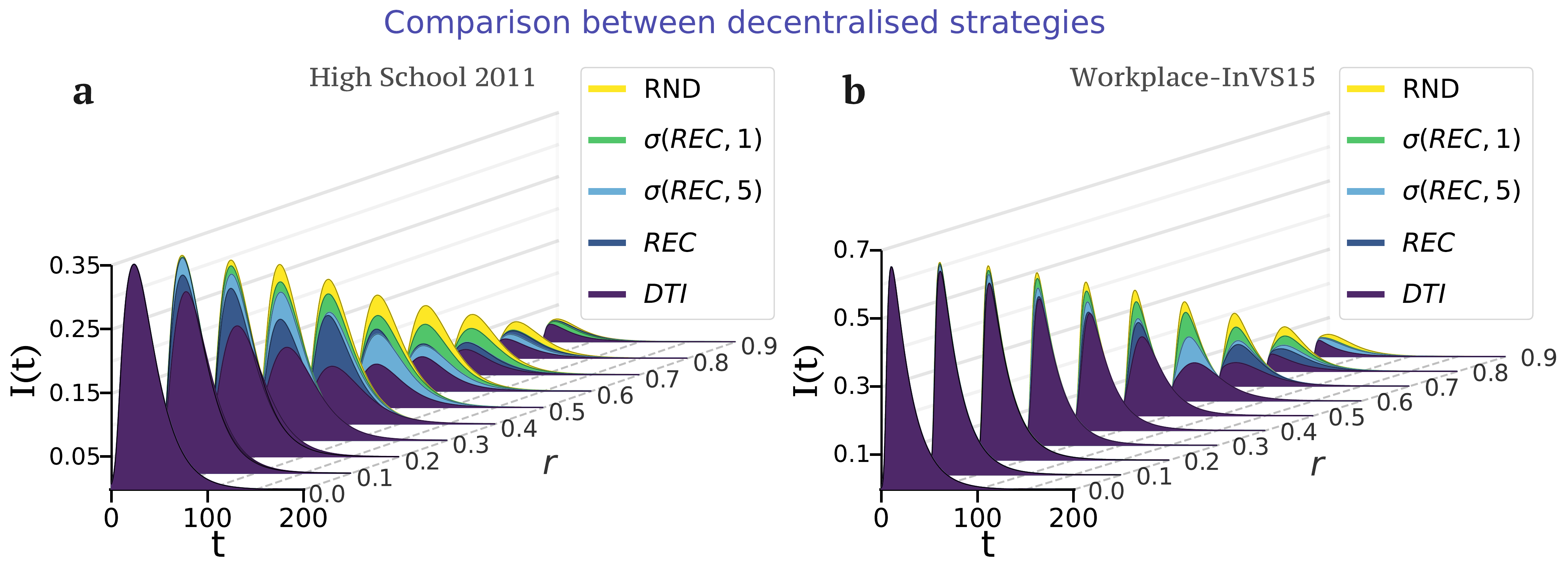}
    \end{center}
    \caption[Mitigation of the epidemic trajectories
    using decentralised CT strategies]{\textbf{Mitigation of the epidemic
            trajectories using decentralised CT strategies}. We report the epidemic
        trajectory when considering different decentralised CT strategies and
        levels of app adoption $r$ for the two real-world systems considered in
        the main text. Overall, the CT decentralised strategies consistently
        lead to a significant reduction of the epidemic trajectory with respect
        to the random case even for small values installation rate $r$, with
        $DTI$ being the best-performing strategy.}
    \label{fig:SI_fig5}
\end{figure*}

\cleardoublepage
\section{SIR model with maximum delay}

The SIR model with the ideal contact tracing presented in the main
manuscript is based on the assumption that any susceptible node with
CT app installed is immediately quarantined (recovered) as soon as one
of their contacts with CT app installed gets infected. However, such
assumption seems too simplistic when applied to real-world systems.
Indeed, people may receive the app notification after several
hours/days from an infectious contact, or worse, they may ignore the app
notification and continue to have contacts with their acquaintances
until they develop symptoms.

As a result, in this section we report the results obtained when
considering a different variant of the SIR model $- SIR_{d} -$ which
accounts for the maximum delay in self-isolation, so that the
neighbours of an infected node only get recovered if an infection
event takes place. In other words, if a node $i$ and its neighbour $j$
have both the contact tracing app installed, $j$ will only become
recovered if $i$ succeeds into infecting $j$. This mimics the fact
that people are in general unwilling to move into self-isolation if
they have no symptoms, and represents the maximum possible delay
between a positive test and self-isolation of it contacts.

To compare the two SIR variants across the 168 real-world networks
analysed in the main manuscript, we report in Supplementary
Fig.~\ref{fig:SI_fig6} the RND and DTI distributions of app
installation rates $r^*$ required to respectively observe a 10, 30,
50\% peak reduction with respect to having no contact tracing in place.
Interestingly, also when considering the $SIR_{d}$ model, the DTI
strategy consistently performs better than the RND. However, all
things being equal, on average we need slightly higher CT adoption
rates $r$ to achieve similar performances in the peak reduction when
considering the $SIR_{d}$ model.

\begin{figure*}[b!]
    \begin{center}
        \includegraphics[width=6.2in]{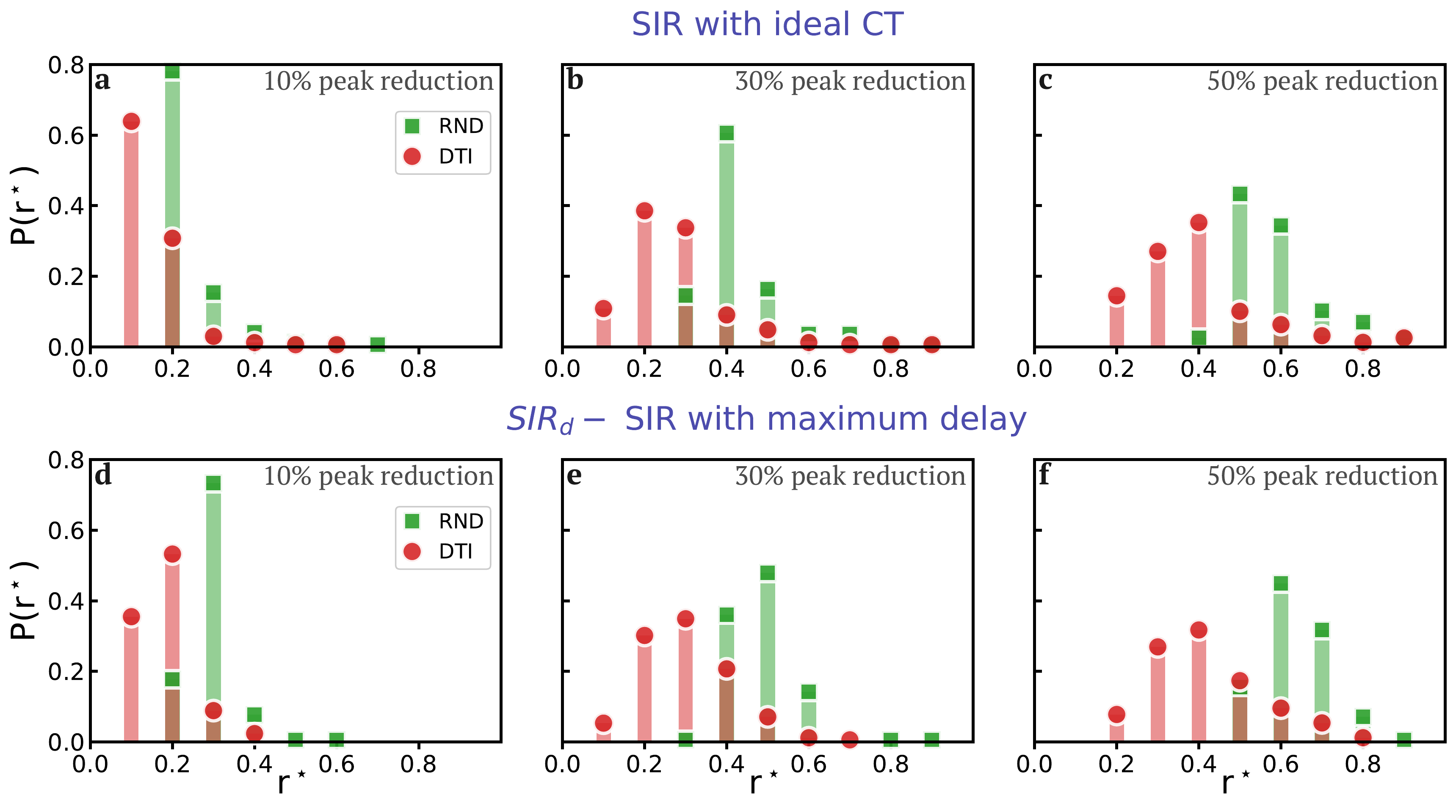}
    \end{center}
    \caption[Performance of CT strategies in mitigating the epidemics
    for the two variants of the SIR model]{\textbf{Performance of CT
            strategies in mitigating the epidemics for the two variants of
            the SIR model}. For the 168 real-world networks analysed,
        we report the distribution of app installation ratios $r^\star$
        required to observe a $10$, $30$, and $50\%$ peak reduction with
        respect to no contact tracing for the DTI (red) and RND (green)
        strategies in the SIR model with ideal contact tracing (a-c)
        and in the SIR model with delay (d-f). Overall, the DTI strategy
        performs far better than the RND as it requires lower values of
        $r$ to induce a similar reduction of the peak in both SIR model
        variants. However, panels (d-f) clearly show that we need slightly
        higher CT adoption rates $r$ to achieve similar performances in
        the peak reduction with the $SIR_{d}$ model.}
    \label{fig:SI_fig6}
\end{figure*}

Furthermore, to easily investigate the impact of the contact tracing
strategies when considering the $SIR_{d}$ model, we ran our simulations
on the same two real-world systems considered in the main manuscript,
the high school and workplace networks. In particular, in Supplementary
Fig.~\ref{fig:SI_fig7} we report the comparison of three different
epidemic indicators for the $SIR_{d}$ model. Overall, the results are
in agreement with the one presented in the main manuscript, so that we
observe a great improvement of the targeted CT strategies with respect
to the random case in all the three epidemic indicators even for low
values of adoption rate $r$. However, we need higher adoption
rates to reach levels of reduction of total number of infected and
infected in the peak similar to the results presented in the main
manuscript, at least 10\% difference of installation rate respectively.

\begin{figure*}[t!]
    \begin{center}
        \includegraphics[width=6.4in]{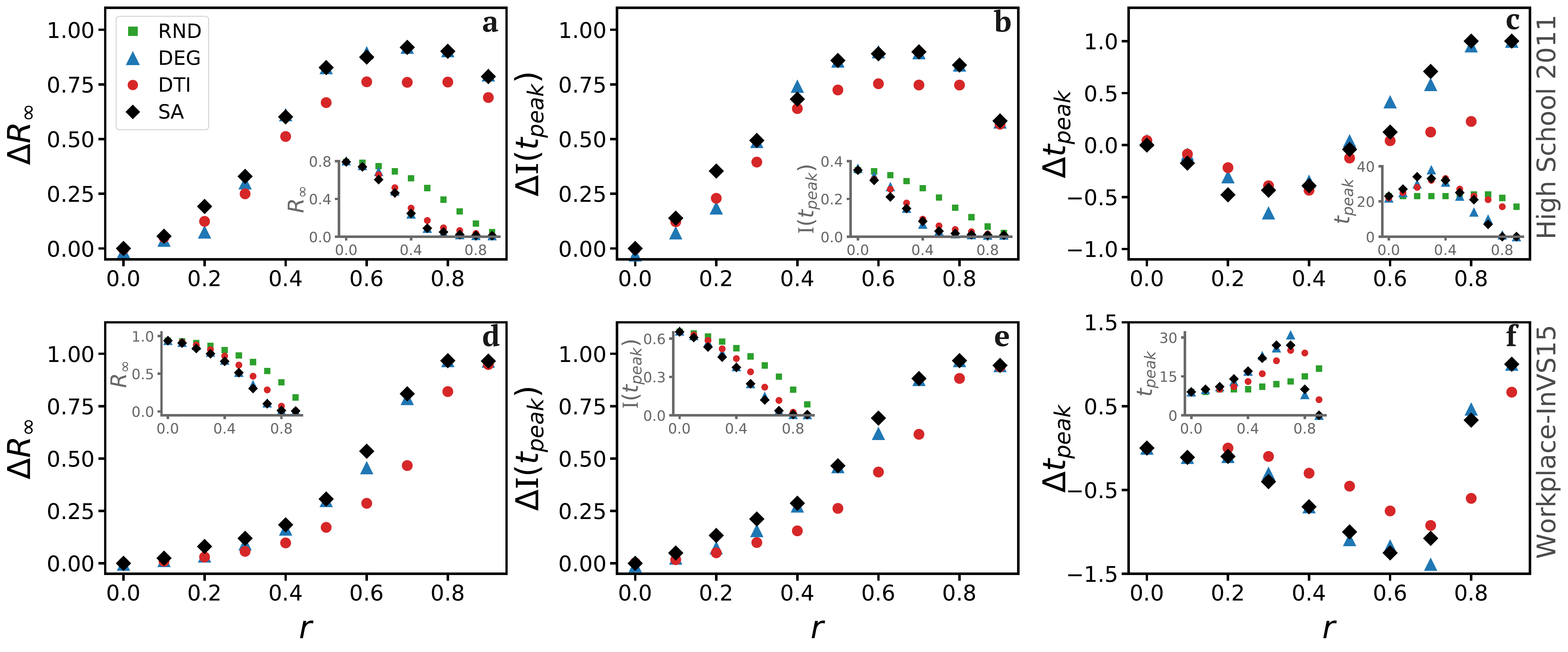}
    \end{center}
    \caption[Comparison of epidemic indicators under different CT
    strategies for the SIR model with delay]{\textbf{Comparison of
            epidemic indicators under different CT strategies for the SIR
            model with delay}. Relative decrease with respect to random
        installations of the total number of recovered $R_{\infty}$,
        height of the infection peak $I(t_{\rm peak})$ and position of
        the peak $t_{\rm peak}$ for DTI, DEG, and SA targeted
        installation when considering the SIR model with delay. For all
        the three synthetic indicators, the targeted strategies provide
        a level of reduction similar to the one presented in the main
        paper, yet achieved with a slightly higher values of
        adoption rate $r$ (10\% increase). The inset of each
        panel reports the plot of the raw variable, respectively
        $R_{\infty}$ (panel a and d), $I(t_{\rm peak})$ (panel b and e)
        and $t_{\rm peak}$ (panel c and f).}
    \label{fig:SI_fig7}
\end{figure*}

Finally, we show that, even in the $SIR_{d}$ model, the decentralised
strategies presented in the main paper (DTI) and in the previous sections
(REC and variants) allow to obtain a substantial reduction of the
epidemic trajectory with respect to the random case. We report in
Supplementary Fig.~\ref{fig:SI_fig8} the comparison between the
decentralised strategies when considering the $SIR_{d}$ model for the
two real-world systems considered in the main manuscript. Remarkably, even
for low values of adoption rate $r$ and also when considering the
maximum delay model $SIR_{d}$, the decentralised CT strategies provide
a consistent reduction in the epidemic trajectory in respect to the
random case.

\begin{figure*}[!b]
    \begin{center}
        \includegraphics[width=6.2in]{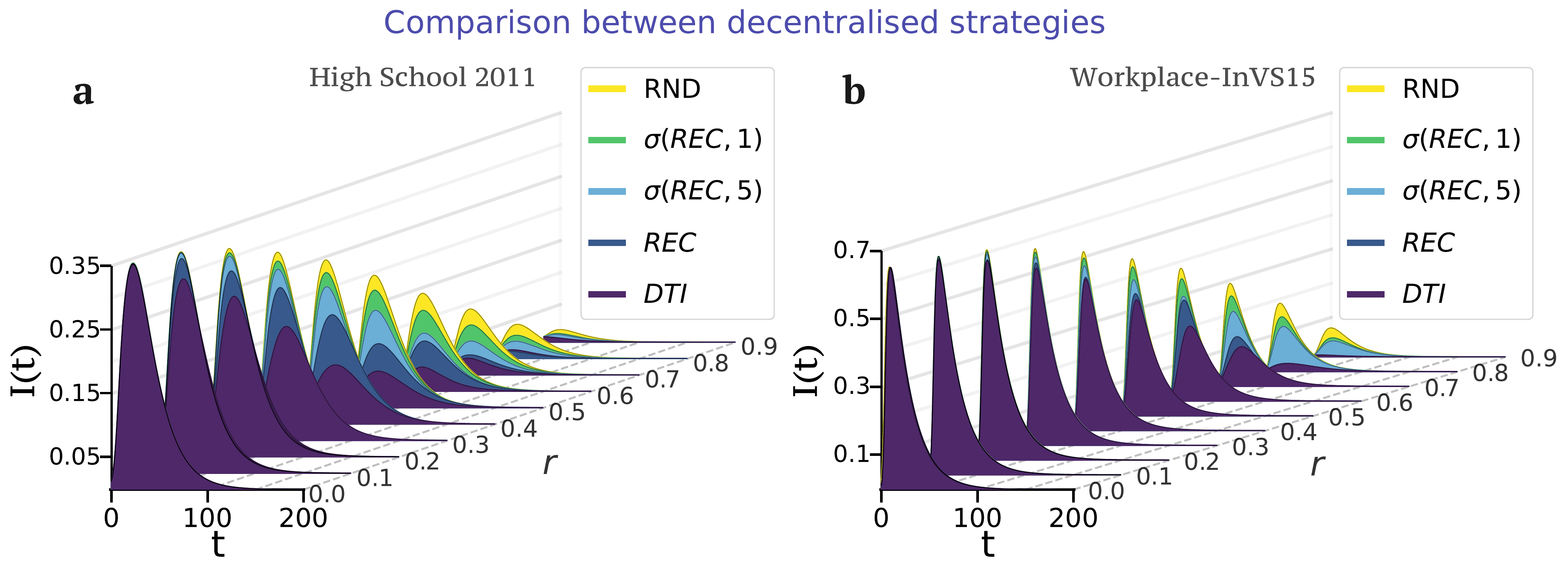}
    \end{center}
    \caption[Mitigation of the epidemic using decentralised strategies
    in the SIR model with delay]{\textbf{Mitigation of the epidemic
            using decentralised strategies in the SIR model with
            delay}. We report the epidemic trajectory when considering
        different CT decentralised strategies and levels of adoption
        rate $r$ for the two real-world systems considered in the main
        manuscript using the $SIR_{d}$ model. Interestingly, the CT
        decentralised strategies consistently lead to a significant
        reduction of the epidemic trajectory with respect to the RND,
        with $DTI$ being the best-performing strategy.}
    \label{fig:SI_fig8}
\end{figure*}

\end{document}